\documentclass[onecolumn,sort&compress,numbers]{els-mrw} % For numbered references Style 

\usepackage{natbib}
\usepackage{hyperref}
\usepackage{amsmath,amssymb,amsfonts,amsthm,makeidx,graphicx}
\usepackage{txfonts}
\usepackage{helvet}
\usepackage{slashed}

\usepackage{tikz-feynman}

\usepackage[outline]{contour}

\usepackage{multirow}

\contourlength{1. pt}

\tikzstyle{loop} = [circle,fill=white,draw=black,minimum size=0.5cm]

\hypersetup{
  colorlinks=true,
  urlcolor=red,
  citecolor=red,
  linkcolor=red,
  filecolor=red
}

\begin{document}

%%%%%%%%%%%%%%%%%%%%%%%%%%%%%%%%%%%%%%%%%%%%%%%%%%%%%%%%%%%%%%%%
%% the following items are mandatory: 
%% - title
%% - author names
%% - affiliation details
%% - abstract
%% - keywords

%% Precise, concise, and informative description of the focus of this work. Avoid abbreviations and formulae in the title
\chapter{Renormalisation }\label{chap1}

%% All author names and affiliations, and email address for corresponding author
\author[1,2]{Leonardo Di Giustino}
%\author[2]{...}%
%
%\author[1,2]{Third Author}%
%
%\address{$^1$,, via Valleggio 11, I-22100, Como, Italy}
%\address{$^2$INFN, Sezione di Milano--Bicocca, 20126 Milano, Italy}

\address[1]{\orgname{University of Insubria}, \orgdiv{Department of Science and High Technology}, \orgaddress{Via Valleggio 11, I-22100, Como, Italy}}
\address[2]{\orgname{INFN}, \orgdiv{Sezione di Milano--Bicocca}, \orgaddress{20126 Milano, Italy}}

\articletag{Chapter Article tagline: update of previous edition, reprint.}

\maketitle

%%%%%%%%%%%%%%%%%%%%%%%%%%%%%%%%%%%%%%%%%%%%%%%%%%%%%%%%%%%%%%%%
%% the following item is mandatory: 
%% 100-150 word summary of the chapter
\begin{abstract}[Abstract]

We give an introduction to renormalisation, focusing first on a pedagogical description of fundamental concepts of the procedure and its features, then we introduce the renormalisation group and its equations. We discuss then the case of gauge theories such as QCD summarising  the current state of the art. We introduce the renormalisation scale setting problem in QCD and we give an illustration of the possible optimisation procedures currently in use.

\end{abstract}

%% 5-10 words that embody the key topics in the chapter. What terms would someone put into a search engine if they were looking for a chapter like this?
\begin{keywords}
 	renormalisation\sep renormalisation group equations \sep QED \sep QCD \sep Standard Model \sep gauge theories \sep scale setting \sep extended renormalisation group 
\end{keywords}

\section*{Objectives}
\begin{itemize}
	\item The renormalisation procedure with its basic concepts and formalism is introduced.	
	\item The renormalisation prescription is applied to QED and QCD.
	\item The renormalisation scheme dependence is discussed.
	\item The renormalisation group and its equations are shown and applied to QCD.
	\item  The renormalisation scale setting problem and its current state of the art is briefly reviewed.
\end{itemize}

%%%%%%%%%%%%%%%%%%%%%%%%%%%%%%%%%%%%%%%%%%%%%%%%%%%%%%%%%%%%%%%%
%% the following items are mandatory: 
%% - Section: Introduction 
%% - further sections
%% - Section: Conclusion
\section{Introduction}\label{intro}

The Standard Model (SM) is theory that gives the most accurate and tested unified description of the strong and electroweak forces up to the large hadron collider (LHC) scales. This theory is based on the local gauge symmetry group $\rm SU(3)_c\times SU(2)_L\times U(1)_Y$ and on the spontaneous symmetry breaking mechanism that has been confirmed by the discovery of the Higgs particle at the LHC. 
Fundamental requirements of the theory are the local gauge symmetry and the {\it renormalizability} Both requirements have to be preserved at any level of accuracy and for any process in the SM.
Renormalisation is a procedure that applies to quantum field theories (QFT) in order to cancel an infinite number of ultraviolet (UV) singularities that arise in loop integration, absorbing them into a finite number of parameters entering the Lagrangian, such as masses, coupling constant and fields. This procedure starts from the assumption that the variables entering the Lagrangian are not the effective quantities measured in experiments, but are unknown functions affected by singularities. The origin of the ultraviolet
singularities is often interpreted as a manifestation that a QFT
is a low-energy effective theory of a more fundamental yet unknown theory.
The use of regularisation UV cut-offs shields the very short
distance domain, where the perturbative approach to QFT ceases to be valid.
Once the coupling has been renormalised to a measured value and at a given energy scale, the effective coupling is no longer
sensitive to the ultraviolet (UV) cut-off nor to any unknown
phenomena arising beyond this scale. Thus, the scale dependence of the coupling can be well understood formally and phenomenologically. This leads to the correct and predictive results of the perturbative calculation.
% Actually gauge theories are affected not only by UV, but also by infrared (IR) divergencies. The cancellation of the latter is guaranteed by the Kinoshita-Lee -Nauenberg (KLN) theorem~\cite{Kinoshita:1962ur,Lee:1964is}.

%%%%
In this chapter we focus on the renormalisation technique and its applications in the framework of the SM, in particular showing applications to QED and QCD. The chapter has a bottom-up structure, starting from fundamental definitions up to the recent state of the art in the QCD renormalisation scale setting problem. This chapter is more intended for those scholars already having some basic knowledge of quantum field theory and who want to quickly become familiar with renormalisation and its recent developments. 
%The chapter is organised as follows:\\
More in detail: in section $\rm I$, we summarise the basic concepts and theoretical foundations underlying the renormalisation procedure; in section $\rm II$ we discuss the case of renormalisation in QED; section $\rm III$ is dedicated to a detailed description of the renormalisation in the QCD; section $\rm IV$ we introduce the scheme dependence and the extended renormalisation group, still referring to QCD; in section $\rm V$ we introduce the renormalisation scale setting problem in QCD and the state of the art in optimisation procedures.

%%%%

%%%%%
\subsection{Power-counting and renormalizability}

 Gauge theories and their relative rules can be derived directly from the quantum fields path integral approach as shown by P. A. M. Dirac, R. P. Feynman and J. S. Schwinger between 1933 and 1951\cite{Dirac:1933xn,Feynman:1948ur,Schwinger:1951xk}. In this formalism one starts from the generating functional $Z$, that corresponds to the integral over all possible paths weighed with their relative phase which is given by the action of the theory:

\begin{align}
Z[J]=\int \mathcal{D} \phi  \, e^{i S[\phi]+i\int d^{\rm d} x J(x) \phi(x)}
\end{align}

where $J(x)$ is the current that generates the field  $\phi(x)$ and $S$ is the action (for an introduction on the functional integral method see Ref. \cite{Zinn-Justin:2002ecy, Itzykson:1980rh}). The action in n.u. (natural units) is dimensionless. The action results from the integral of a local Lagrangian density in the $\rm d$-dimensions space-time:

\begin{align}
S=\int \mathrm{d}^{\mathrm{d}} x {\cal L}(x)
\end{align}

 it follows that the Lagrangian density has mass dimension:

\begin{align}
[{\cal L}(x)]=\mathrm{d}
\end{align}

and is the sum of non-interacting ($\rm free$) fields terms and interacting ($\rm I$) terms:

\begin{align}
{\cal L}(x)=\sum_i {\cal L^{\rm i}_{ \rm free}}+\sum_i {\cal L^{\rm i}_{\rm I}}
\end{align}

General forms of the fermion, scalar and gauge field kinetic terms are :

\begin{align}
S=\int \mathrm{d}^{\mathrm{d}} x \bar{\psi} i   \slashed{\partial} \psi, \quad S=\int \mathrm{d}^{\mathrm{d}} x \frac{1}{2} \partial_\mu \phi \partial^\mu \phi ,  \quad S=\int \mathrm{d}^{\mathrm{d}} x \frac{1}{4} X_{\mu \nu} X^{\mu \nu}.
\label{eq:kinterms}
\end{align}

from Eq. \ref{eq:kinterms} follows that the dimensions of fermion, scalar and gauge fields are:

\begin{align}
[\psi]=\frac{1}{2}(d-1), \quad[\phi]=\frac{1}{2}(d-2),  \quad[A_{\mu}]=\frac{1}{2}(d-2)
\end{align}

where the gauge field strength $X_{\mu \nu}=\partial_\mu A_\nu-\partial_\nu A_\mu$ has a single derivative of $A_\mu$, so $A_\mu$ has the same dimension as a scalar field. 

The coupling between matter and gauge fields in gauge-theories is introduced by the {\it minimal gauge coupling} (see Ref. \cite{Peskin:1995ev}) in the covariant derivative : 
\begin{align} 
D_\mu=\partial_\mu-i g T_a A^a_\mu,
\end{align} where $T^a$ are the SU(N) gauge group generators (see the Chapter on the Group Theory) and according to the convention repeated indices are summed, given that both terms have the same dimension, we have that for $d=4$, the dimension of the minimal gauge coupling is :

\begin{align*}
\left[D_\mu\right]=1 \quad  \Rightarrow \quad \left[g \right]=0
\end{align*}

This result leads to classify the gauge theories as {\it renormalizable theories}. In general more complex structures of the interaction operators in the Lagrangian may arise, especially in effective field theories (EFT), and in order to classify them, a more general approach is introduced, namely the {\it power-counting}. This method is fundamental to determine the {\it superficial degree of divergence} of a diagram or the dimension of coupling.
For example the number of loops $L$ in a given diagram can be determined by :
\begin{align}
L=(Bp)+(Fp)-n+1,
\end{align}
where $Bp$ and $Fp$ are the number of internal lines in the graph, $n_i$ is the number of vertices of the $ i^{\rm th}$-type,  $n-1\equiv \sum_i n_i-1$ is the total number of vertices except one which is subtracted out because of the overall momentum conservation.
Considering that in 4 dimensions each loop brings a power 4 in the integration and that a fermion propagator has carries a power of $p^{-1}$, while the scalar propagator $p^{-2}$ and that at each vertex depending on the type of interaction there might be $d_i$ number of derivatives, the superficial degree of divergence results:

\begin{align}
\label{eq:Dsupdeg}
D_s=4 L-2(Bp)-(Fp)+\sum_i n_i d_i
\end{align}

One can relate fermion and scalar lines as follows:
Counting the scalar and fermion lines, we get

\begin{align}
Be+2(Bp) &=\sum_i n_i B_i \label{eq:Bcons} \\
Fe+2(Fp)&=\sum_i n_i F_i
\label{eq:Fcons}
\end{align}
where the $Be,Fe$ are the number of external scalar and fermion lines in a graph , while 
$B_i,F_i$ are the number of scalar and fermion lines entering in each vertex of  $i^{\rm th}$-type.
Using Eqs.\ref{eq:Bcons}, \ref{eq:Fcons} in Eq. \ref{eq:Dsupdeg} , we obtain:

\begin{align}
D_s=4-Be-\frac{3}{2} Fe+\sum_i n_i \delta_i
\end{align}

where

\begin{align}
\delta_i=B_i+\frac{3}{2} F_i+d_i-4
\end{align}

is called the {\it index of divergence of the interaction}. Considering that in 4-dimensions the Lagrangian density $\mathcal{L}$ has dimension 4 and scalar field, fermion field and the derivative have dimensions, $1, \frac{3}{2}$, and 1 respectively, we obtain for the dimension of the coupling constant $g_i$ the following result:

\begin{align}
\left[g_i\right]=4-B_i-\frac{3}{2} F_i-d_i,
\end{align}
it follows that $\left[g_i\right]=-\delta_i$.
According to the coupling dimension we can distinguish 3 different cases:
\begin{enumerate}
\item   $\delta_i<0 $, In this case, $D_s$ decreases with the number of $i^{\rm th}$-type of vertices. Thus we have only a finite number of divergent diagrams at lower orders and
this interaction leads to a {\it super--renormalizable } theory;
\item  $\delta_i=0 $, $D_s$ is independent of the number of $i^{\rm th}$-type vertices. Divergences arise at all orders in perturbation theory but in a finite number of Green’s functions. These interactions lead to $renormalizable$ theories;
\item   $\delta_i>0 $, $D_s$ increases with the number of $ i^{\rm th}$-type vertices and an infinite set of divergent Green’s functions arises in perturbation theory. These type of interactions leads to the so called {\it non--renormalizable} theories.
\end{enumerate}
%%%%%%%%%%%%%%%%%%%%%%%%%%%%%%%%%%%%%%%%%%%
\subsection{Divergences and regularisation}
At 1-loop accuracy, the superficial degree of divergence $D_s\geq 0$ determines the degree of divergence of a Feynman diagram, as shown in the previous section, while  $D_s<0$ determines the convergence of a Feynman diagram. Going beyond the 1-loop accuracy there might also occur other cases where the overall $D_s$ is negative but some subgraphs might be divergent. By using the Weinberg’s theorem \cite{Weinberg:1959nj} we can single out without ambiguity convergent from divergent diagrams. According to the Weinberg’s theorem a Feynman diagram is convergent if all superficial degree of divergence $D_s$ of the graph and of each possible subgraph in it, are all negative.
In general we can distinguish 4 cases:
the case of {\it primitively divergent graph} when the overall superficial degree of divergence is nonnegative but all subgraphs are convergent; 
then we have the case of {\it disjoint divergences} when we have 2 or more disjointed subgraphs that are divergent; another case is that of {\it nested divergences} when 2 or more divergent 1PI diagrams are completely contained one another and cannot be separated;  and the last case that does not match any of the previous cases is the case of {\it overlapping divergences}, ( for further discussion on this topic we remind to Refs. \cite{Collins:1984xc,Cheng:1984vwu}). 
In order to determine the correct asymptotic behavior of a Feynman diagram and consequently to introduce the renormalisation procedure, it is necessary to introduce a regularizing UV-cutoff ($\Lambda_{\rm UV}$), which controls the asymptotic limit $\Lambda_{\rm UV}\rightarrow \infty$ of the graph.
Several regularisation prescriptions exist: the {\it covariant regularisation} (Pauli and Villars\cite{Pauli:1949zm}) in which a cutoff regulator is introduced by means of a redefinition of the Feynman propagator, the {\it lattice} in which space-time is discretized and the lattice spacing acts as a natural UV-regulator $\Lambda_{\rm UV}\sim \hbar/a$( K. Wilson \cite{Wilson:1974sk,Wilson:1975hf}); 
and then the dimensional regularisation introduced in the chapter on {\it perturbative QCD} by Gudrun Heinrich {\it et al.}. In this
procedure~\cite{tHooft:1972tcz,Ashmore:1972uj,Cicuta:1972jf,Bollini:1972ui} one varies the dimension of the loop integration as $D=4-2\varepsilon$,
and introduces a scale $\mu$ in order to restore the correct
dimensionality of the coupling and UV-divergences occur as poles in the infinitesimal dimension variation ${\varepsilon^{-1}}$.
In this chapter we will refer to dimensional regularisation in order to introduce the renormalisation procedure in particular to QCD and to the Standard Model.

%%%%%%%
\subsection{The renormalisation prescription}

Renormalisation is a procedure made of a sequence of mathematical passages aiming to consistently isolate and remove all the UV-singularities arising in loop integration.  This procedure is crucial for the development
of a relativistic quantum field theory and together with the gauge invariance is the basic requirement for developing a reliable theory for describing the fundamental interactions of elementary particles entering the SM. \\
By renormalisation, UV-singularities are cast into the redefinition of the parameters entering the initial Lagrangian that are commonly indicated as  {\it bare parameters}  and  {\it bare} Lagrangian. This stems from the basic idea that the parameters of the bare-Lagrangian are not those physically measured and are also unknown divergent functions. Once UV-singularities are cancelled from the parameters these become {\it renormalized-parameters}, they are no longer sensitive to the UV-cutoff and perturbatively “converge” to the measured physical quantities. In fact, the renormalisation prescription acts order-by-order in perturbation theory where new inifinities arise and must be reabsorbed.
The renormalized Lagrangian leads to a theory which is finite and predictive all over the spectrum of the accessible physical energies.
Given that a gauge-theories are renormalizable, a finite set of Green’s functions only must be renormalized.
It is a common practice to apply renormalisation first to Green’s functions related to Feynman diagrams that cannot be disconnected by a single cut of an internal line (propagator), namely the one-particle-irreducible (1PI) diagrams.

The other way, i.e. to apply renormalisation directly to connected Green’s functions, is also possible but certainly cumbersome and since any connected diagram can be decomposed into 1PI diagrams, as shown in Fig. \ref{fig:1pidiag}, without introducing further loops it is also not necessary. 
\begin{figure}[h]
\begin{align}
 \begin{tikzpicture}
 \node at (0,0) { a.};
 \end{tikzpicture} 
  \hspace{0.9cm} & \begin{tikzpicture}
\begin{feynman}[small]
\node  (a) at (0.2,0);
\node [blob, baseline = (a.base)] (b) at (1.7,0);
\node (c) at (3.2, 0);
\node  at (3.4 ,0){ $=$};
\diagram* {(a) -- (b) -- (c)} ; 
\node (k) at (3.7,0);
\node (j) at (5.5,0);
\node at (5.6 ,0) {\contour{white}{\scriptsize \bf $+$}};
\node (e) at (5.7,0) ;
\node [blob] (f) at (7,0) {\contour{white}{\scriptsize \bf 1PI}};
\node (h) at (8.3,0);
\node at (8.4 ,0) {\contour{white}{\scriptsize \bf $+$}};
\node (g) at (8.5,0);
\node [blob] (i) at (9.8,0) {\contour{white}{\scriptsize \bf 1PI}};
\node [blob] (l) at (10.8,0) {\contour{white}{\scriptsize \bf 1PI}};
\node (m) at (12.1,0);
\node at (12.7 ,0) { $+$ \,  {  $\cdots$}};
\diagram* {(k)--(j), (e) -- (f) -- (h), (g)--(i)--(l)--(m)} ; 
\end{feynman} 
\end{tikzpicture} \nonumber\\
b. \hspace{1cm} &  \frac{1}{A-B} =\frac{1}{A}+\frac{1}{A}{  (B)} \frac{1}{A}+\frac{1}{A}(B)\frac{1}{A}(B)\frac{1}{A}+\cdots \nonumber \\
c.  \hspace{1cm}  &  \frac{i}{p^2-m^2_0-G^P(p^2)+i\varepsilon} = \frac{i}{p^2-m^2_0+i\varepsilon}+\frac{i}{p^2-m^2_0+i\varepsilon} \left(-i G^P(p^2) \right)  \frac{i}{p^2-m^2_0+i\varepsilon}+\cdots \nonumber 
\end{align} 
	\caption{{\bf  a)}1PI-diagram contributions to the Feynman propagator; {\bf  b)} summed geometric series with A the propagator at lowest order and B  the1PI insertion; {\bf  c)} exact scalar propagator in momentum space.}
	\label{fig:1pidiag}
\end{figure}

%%%%
According the {\it  conventional renormalisation}\cite{Peskin:1995ev,Collins:1984xc,Cheng:1984vwu}, one starts with the bare Lagrangian and order-by-order cancels the infinities arising in the amputated Green’s function by the redefinition of the parameters, i.e. matter and gauge fields, masses and the coupling. 
Following this procedure, it is necessary first to regularize the integral, e.g.
by using the dimensional regularisation in which the loop integrals become functions in the $\varepsilon$ dimension regulator and diverge in the zero limit: $\lim_{\varepsilon\rightarrow 0} \varepsilon^{-1} \rightarrow \infty$.
Starting from the 1-loop integral, once the integral is regularized, one has to Taylor expand the Green’s function in the external momentum squared, or any Lorentz-invariant ( such as $s,t,u,...$), that for sake of simplicity we define as $p^2$, around an arbitrary {\it subtraction point} $\mu^2$:
\begin{align}\label{eq:taylor}
G \left(p^2\right)=a_0+a_1 (p^2-\mu^2) +\ldots \frac{1}{n!} a_n\left(p^2-\mu^2\right)^n+\ldots,
\end{align}
where :
\begin{align}
a_n=\left.\frac{\partial^n}{\partial^n p^2} G \left(p^2\right)\right|_{p^2=\mu^2}
\end{align}
and only the first few derivatives are singular according to the superficial degree of divergence of the graph, e.g. in case $D_s=0$ only $a_0$ is logarithmically divergent, while if $D_s=2$, we have that $a_0$ and $a_1$ are quadratically and logarithmically divergent respectively\footnote{The term linear in $p^\mu$ is missing because it is not Lorentz invariant.}.
Thus we can write Eq \ref{eq:taylor} according to the superficial degree of divergence $D_s=0$ or $D_s=2$ respectively, separating divergent and finite parts as :
\begin{align}
G^V (p^2)=G^g(\mu^2)+\tilde{G^V}(p^2),\\
G^P (p^2)=G^{m}(\mu^2)+G_\phi^\prime (\mu^2)(p^2-\mu^2)+\tilde{G}^P(p^2),
\end{align}
where $G^g(\mu^2)$, $G^m(\mu^2)$, $G_\phi^\prime (\mu^2)$ are the singular terms that cancel the divergences related to $g_0,m_0,\phi_0$, while $\tilde{G^{V,P}}(p^2)$ are the finite parts of the vertex (labelled with superscript “V”) and the propagator (labelled with superscript “P”) and they obey the {\it renormalisation scheme} conditions: 
\begin{align}
\tilde{G^V}(\mu^2)=0,
\end{align}
for $D_s=0$ and both:
\begin{align}
\tilde{G^P}(\mu^2)=0,\\
\tilde{G^P}^\prime(\mu^2)=0,
\end{align}
for $D_s=2$. This scheme is known as {\it momentum subtraction scheme} (MOM).
Thus, in order to subtract the infinite parts of the integrals, renormalisation conditions must apply. 

The parameters appearing in the lowest order of calculation $g_0, m_0, \phi_0$ are renormalized introducing a renormalisation constant $Z_i$,with $i=g, m, \phi,$ which subtracts the corresponding divergent contribution arising from the 1-loop calculation:
%%%%%%%%%%%%%%%%%%%%%%%

\begin{align}\label{eq:rencond1}
& \phi  =Z_\phi^{-1 / 2} \phi_0,\\
& g  =Z_{g}^{-1} Z_\phi^2 g_0, \label{eq:rencond2} \\
& m^2  =m_0^2+\delta m^2, \label{eq:rencond3}
\end{align}
where the renormalisation constants are:
\begin{align}\label{eq:zconv}
&Z_\phi= 1+G_\phi^\prime (\mu^2),  \\
& Z_{g}= 1+G^g(\mu^2),\\
&\delta m^2=G^{m}(\mu^2).
\end{align}
Once the 1PI graphs have been renormalized, all Green’s functions become finite and the dependence on the UV-regulator is removed.
Thus any renormalized  $n$-point Green's function when we express the bare mass $m_0$ and bare coupling constant $g_0$ in terms of the renormalized mass $m$ and coupling $g$, and multiply by $Z_\phi^{-1 / 2}$ for each external field as in Eq. \ref{eq:rencond1},  the renormalized $n$-point Green's function becomes:

\begin{align}
\label{eq:npointgreen}
G_{\mathrm{R}}^{(n)}\left(p_1, \ldots, p_n ; g, m\right)=Z_\phi^{-n / 2} G_0^{(n)}\left(p_1, \ldots, p_n ; g_0, m_0, \varepsilon \right)
\end{align}

where $\varepsilon$ is the UV-regulator. 
From Eq. \ref{eq:npointgreen} it follows the property that the n-point Green’s functions are {\it multiplicatively renormalizable}, i.e. once the parameters have been replaced with the renormalized parameters the divergences 
are cast into the multiplicative constants $Z_\phi^{-n / 2}$.
Amputated Green’s functions $\Gamma^{(n)}_R(p_1,\ldots,p_n)$ are related to the renormalized n-point Green’s function by:
\begin{align}
G_{\mathrm{R}}^{(n)}\left(p_1 \ldots p_n\right) & =\prod_{j=1}^n\left[\mathrm{i} \Delta_{\mathrm{R}}\left(p_j\right)\right] \Gamma_{\mathrm{R}}^{(n)}\left(p_1 \ldots p_n\right)
\end{align}
which stems from the definition:
\begin{align}
G_{\mathrm{0}}^{(n)}\left(p_1 \ldots p_n\right) & =\prod_{j=1}^n\left[\mathrm{i} \Delta_{\mathrm{0}}\left(p_j\right)\right] \Gamma_{\mathrm{0}}^{(n)}\left(p_1 \ldots p_n\right)
\end{align}

where $\Delta_{\mathrm{R}}(p_j)=Z^{-1}_{\phi}\Delta_0(p_j)$ is the renormalized Feynman propagator for each external particle. 

Thus, amputated Green's functions are made finite by replacing the bare quantities $g_0, m_0$ with the physical quantities $g, m$,  and multiplying by the renormalisation constant $Z_\phi^{n / 2}$,i.e.: 

\begin{align}
\Gamma_{\mathrm{R}}^{(n)}\left(p_1, \ldots, p_n ; g, m \right)=Z_\phi^{n / 2} \Gamma_0^{(n)}\left(p_1, \ldots, p_n ; g_0, m_0, \varepsilon \right) .
\label{eq:gammar}
\end{align}

\begin{figure}[h]
\begin{align}
\begin{tikzpicture}
\begin{feynman}[small]
\node (a) at (-2,-2);
\node (s) at (-3,0){\contour{white}{ \bf $G^{(4)}=$ }};
\node [circle,fill=lightgray,draw=black, minimum size=1.3cm](b) at (0,0){\contour{white}{ \bf $\Gamma^{(4)}$ }};
\node (c) at (2, 2);
\node (d) at (-2, 2);
\node (e) at (2, -2);
\node [blob,minimum size=0.6cm] (f) at (1.2,1.2) {\contour{white}{\scriptsize \bf }};
\node [blob, minimum size=0.6cm] (i) at (-1.2,1.2) {\contour{white}{\scriptsize \bf }};
\node [blob,minimum size=0.6cm] (k) at (-1.2,-1.2) {\contour{white}{\scriptsize \bf }};
\node [blob,minimum size=0.6cm] (l) at (1.2,-1.2) {\contour{white}{\scriptsize \bf }};
\diagram* {(a)--(k)-- (b) --( f)--(c), (d)--(i)-- (b) --(l)--(e)} ; 
\end{feynman} 
\end{tikzpicture} \nonumber
\end{align} 
	\caption{Structure of the 4-points connected Green’s function $G^{(4)}$, decomposed into exact propagators and amputated $\Gamma^{(4)}$ Green’s function.}
	\label{fig:amputated}
\end{figure}
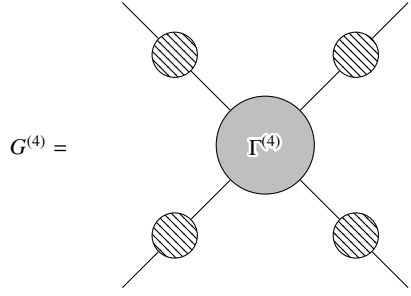

%%%%

\subsection{Scale invariance: the RGE}

%%%%%%%%%%%%%%%%%%%%%%%%%%
We have previously shown how renormalisation conditions, Eqs. \ref{eq:rencond1} to \ref{eq:rencond3},  introduce the scale and scheme dependence into the Lagrangian, by fixing the subtraction point $\mu^2$ and the finite part of the renormalized Green’s functions $\tilde{G}(\mu^2)$.
Given that the original bare-Lagrangian is scale and scheme invariant and that any prediction cannot depend on a conventional choice, we must recover scale and scheme invariance in a theory. The latter is obtained by requiring observables being scale and scheme invariant under the {\it renormalisation group equations} (RGE).
The renormalisation group was first introduced by Stueckelberg and Peterman in 1953 \cite{Stueckelberg:1952hma} and then applied by Gell-Mann and Low to QED in 1954\cite{Gell-Mann:1954yli}.
There are several ways to set up the renormalisation
group equations, in this section we refer to the Callan-Symanzik
equation~\cite{Callan:1970yg,Symanzik:1971vw} .

Let us consider for example the case of the ratio:
 \begin{align}
 \label{eqn:R}
 R_{e^+
e^-}
(s)=\frac{\sigma(e^+e^-\rightarrow\mbox{hadrons})}{\sigma(e^+e^-\rightarrow
\mu^+ \mu^-)}
 \end{align}
at high energy , $s>> m_f^2$, and thus neglecting the masses. The two cross sections at the numerator and denominator, are given at the lowest order by:
  \begin{align}
 \sigma(e^+e^-\rightarrow
{\rm hadrons })=\frac{4\pi\alpha^2}{3s}N_c\sum_f Q_f^2,
 \end{align}
 and 
  \begin{align}
\sigma(e^+e^-\rightarrow \mu^+ \mu^-)=\frac{4\pi\alpha^2}{3s},
 \end{align}
respectively, with $\alpha\equiv e^2/4\pi$ is the QED coupling constant.
As shown in the formula the single cross sections depend on the
center-of-mass energy $s=Q^2$, but this dependence cancels in the
$R_{e^+ e^-}$ at lowest order. Also by dimensional analysis, a
constant value of the observable $R_{e^+ e^-}$ would be predicted
independently of any energy scale $s$, the ratio being a
dimensionless quantity. However, higher order loop integrations
and the renormalisation procedure of the coupling constant
introduce a scale dependence. Since $R$ is dimensionless and since masses are neglected in first approximation, the scale dependence is
introduced only via the scale dependence of the strong coupling
$\alpha_{s}(\mu)$ and through a ratio
$\left(Q^{2}/\mu^{2}\right)$-like dependence of the perturbative
coefficients. In fact, except for the first two terms
$r_{0},r_{1},$ that are scale independent, the coefficients
$r_{n}$ are polynomials of $\ln\left( Q^{2}/\mu^{2}\right)$ with
highest power $n-1$. By means of the RGE all these logarithms can
be reabsorbed into the running coupling. The purpose of taking
$\alpha_{s}$ scale-dependent is to transfer to $\alpha_{s}$ all
terms involving $\mu$ in the perturbative series of
$R\left(s\right)$.
 The independence of $R$ with respect to $\mu$ is given by the Callan-Symanzik
relation for QCD:

 \begin{align}
 R_{e^+ e^-}(s;\mu)=N_c\sum_f Q_f^2\left[1+R(s;\mu)\right].
 \end{align}
where $Q_f$ are quark charges summed over the flavor index $f$ and
\begin{align}R(s;\mu)=\sum_{n=1}^{\tilde{n}} r_n(s;\mu) \left(
\alpha_s(\mu)/\pi\right)^n\end{align}.

The scale invariance for the observable $R_{e^+e^-}$, is given by the total derivative in the scale:

\begin{align}\mu^2 \frac{d}{d\mu^2}R_{e^+ e^-}(s,\mu)=0, \end{align} or equivalently:
\begin{align}\label{eqn:Rcallan} \left[ \mu^2\frac{\partial}{\partial
\mu^2}+\beta(\alpha_s)\frac{\partial}{\partial \alpha_s}\right]
R_{e^+ e^-}(s;\mu)=0, \end{align} where \begin{align}\label{eqn:betag}
\beta(\alpha_s)=\mu^2 \frac{\partial \alpha_s}{\partial \mu^2},
\end{align} 
is known as $\beta$-function and it governs the evolution of the strong coupling with the scale.
By setting the renormalisation scale $\mu$ equal to the physical
scale $Q^2=s$ would remove the $\ln(Q^{2}/\mu^{2})$ in the
coefficients $r_{n}$ and fold the $\mu$-dependence into
$\alpha_{s}\left(\mu^{2}=s\right)$. Thus the option of choosing
$\mu^{2}=s$ yields the simplest form for the perturbative
expansions of given observable.
Relations that are analogous to Eq. \ref{eqn:Rcallan} and that involve also  masses, exist for any type of renormalized quantity or correlation function.
In general an amputated renormalized Green’s function depends on the renormalized parameter $g, m$, and on the subtraction point $\mu$. From Eq. \ref{eq:gammar}, we can derive the {\it renormalisation group equations} for the renormalized n-point amputated Green’s functions:

\begin{align}
\left[ \mu^2 \frac{\partial}{\partial
\mu^2}+\beta(g)\frac{\partial}{\partial g}-n \gamma+  \gamma_m  m \frac{\partial}{\partial m}   \right]
\Gamma_{\mathrm{R}}^{(n)}\left(p_1, \ldots, p_n ; g, m ,\mu \right)=0
\label{eq:callan1}
\end{align}
where:
\begin{align}
\gamma & =\mu^2 \frac{\partial}{\partial \mu^2} \log Z^{1/2}_{\phi}, \label{eq:anomalousphi} \\
\gamma_m & = \mu^2 \frac{\partial}{\partial \mu^2} \log m \label{eq:anomalousm},
\end{align}
are the {\it anomalous dimensions} for the field and the mass respectively. 
These are pertubatively calculates series in the coupling $g$ and they are responsible for a modified behavior of a renormalized quantity in the asymptotic limit, i.e. at very large values of the renormalisation scale $\mu$\footnote{The asymptotic limit is also referred to as {\it deep Euclidean region}, since the two regions are mapped one another by the  {\it Wick rotation}(see Ref. \cite{Cheng:1984vwu})}. 
In fact, at large scales, we would expect that 1PI-Green’s function would scale according to the Weinberg's theorem\cite{Weinberg:1959nj} which states that: for non exceptional momenta parametrised as $p_i=\sigma k_i$, the amputated Green's function $\Gamma_R^{(n)}$ in the deep Euclidean region (corresponding to $\sigma\rightarrow \infty$ with $k_i$ fixed ), scales as $\sigma^{4-n}$ times a polynomial in $\ln{\sigma}$ of finite order in the coupling $g$. It is noted that Green’s functions scale with a power which is given by their superficial degree of divergence. The anomalous dimension $\gamma$ corrects this behavior, since the logarithmic terms in the polynomials, may recursively resum at all orders in perturbation theory, inducing an additional exponent in the scaling behavior: $\sim \sigma^{4-n-\gamma}$ (see e.g. Ref. \cite{Peskin:1995ev}).

Hence, RGE give a quantitatively description of how an overall shift of the scale  $\mu$ in a correlation function $\Gamma_R^{(n)}\left(p_1, \ldots, p_n ; g, m, \mu\right)$, is exactly compensated by the variation of all renormalized quantities, such as $g=g(\mu)$, $m=m(\mu)$ and $Z_{\phi}(\mu)$.

%%%%
\section{BPHZ renormalisation and QED}

An equivalent renormalisation prescription has been developed by Bogoliubov, Parasiuk, Hepp and Zimmermann \cite{Bogoliubov:1957gp,Hepp:1966eg,Deser:1970spa}, known as BPHZ renormalisation. \\
The BPHZ renormalisation is organized differently from the conventional renormalisation:
\begin{enumerate}
\item One starts directly with the renormalized Lagrangian ${\cal L}$ to derive the Feynman rules of the theory.
\item  Thus, singularities are singled out by separating the divergent parts of 1PI diagrams by Taylor expansion. 
\item Subsequently, a set of {\it counterterms} $\Delta \mathcal{L}^{(1)}$  is designed and introduced in the $\cal L$ to cancel the 1-loop divergences.
\item The Lagrangian corrected at 1-loop : $\mathcal{L}^{(1)}=\mathcal{L}+\Delta \mathcal{L}^{(1)}$ is then used to generate the higher 2-loop corrections iterating the procedure at 2-loops accuracy and determine the counterterms $\Delta \mathcal{L}^{(2)}$ that cancel the 2-loop divergences and so on.
\item By iterating the procedure at all orders, one obtains the final Lagrangain:

\begin{align}
\mathcal{L}_{f}=\mathcal{L}+\Delta \mathcal{ L}
\end{align}
where the counterterm Lagrangian $\Delta \cal{L}$ is given by,
\begin{align}
\Delta \mathcal{L}=\Delta \mathcal{L}^{(1)}+\Delta \mathcal{L}^{(2)}+\cdots \Delta \mathcal{L}^{(n)}+\cdots
\end{align}

\end{enumerate}

We show this procedure by applying to QED,  the renormalized QED lagrangian is:

\begin{align} 
\label{eq:Lqed}
\mathcal{L}_R^{\rm QED}= -\frac{1}{4} F_{\mu \nu} F^{\mu \nu}+\bar{\psi}(i \slashed{\partial}-m) \psi-e \bar{\psi} \gamma^\mu \psi A_\mu 
\end{align}

Renormalized fields are related to the bare-fields by:

\begin{align}
\label{eq:qedz2}
\psi & \equiv Z_2^{-1 / 2} \psi_0 \\
\label{eq:qedz3}
A_\mu & \equiv Z_3^{-1 / 2} A^0_\mu
\end{align}
where $Z_2$ and $Z_3$ are ther renormalisation constants resulting from the 1-loop corrections to the Feynman propagators for the fermions and the photon respectively.

The coupling and mass renormalisation are introduced to cancel the remaing divergences from the vertex and fermion propagator respectively, by the following relations:

\begin{align}
\label{eq:qedcoupling}
e Z_1 \equiv e_0 Z_2 Z_3^{1 / 2}
\end{align}
and:
\begin{align}\label{eq:qedmass}
m+\delta m \equiv Z_2 m_0
\end{align}

The renormalisation condition Eq.~\ref{eq:qedcoupling} sets the renormalized coupling, $e$, at a given momentum $\mu^2$ known as subtraction point or {\it renormalisation scale}, to the value obtained from a precise experimental measurement. In the BPHZ renormalisation the value of $\mu^2=0$.

In particular, as shown by Gell-Mann and Low~\cite{Gell-Mann:1954yli}, the scale dependence of the QED coupling is well described by the effective coupling that in the $\overline{\textrm{MS}}$ scheme has the analytic formula:
\begin{equation}
{\alpha(Q)} = {\alpha_0 \over  {\left(1 - \Re e
\Pi^{\overline{\textrm{MS}}} (Q^2)\right)}},
\end{equation}
where the vacuum
polarization function ($\Pi$) is perturbatively calculated including contributions from leptons, quarks and gauge-bosons in the loop, while the renormalized value of the QED {\it fine structure constant} is set to: $\alpha_0^{-1}=\left(\left.\frac{g^2(q^2)}{4 \pi}\right|_{q^2=0}\right)^{-1} =137.036$. 
At the same subtraction point also the values of the finite parts of the loop integrals are fixed and these values define the {\it renormalisation scheme}.
The same values of the renormalisation scale and scheme are also common to the other renormalisation conditions: i.e. Eqs. \ref{eq:qedz2}, \ref{eq:qedz3} and \ref{eq:qedmass}.
These two operations are not free from ambiguities, in fact both the subtraction point and the finite part are arbitrary. One may decide to make the subtraction at a different value of the scale $\mu^2$ and defining a different scheme , i.e. choosing another value of the finite term by subtracting out, together with the divergent term not only the pole but also an extra finite constant. This arbitrariness  leads to ambiguities that need to be fixed in order to make reliable theoretical predictions and they will be discussed in the following sections.

The counterterms entering in the  $\Delta \mathcal{L}^{(1)}$ are defined by considering the small perturbations $\delta_i$ arising from the radiative loop corrections:

\begin{align}
\delta_1 &\equiv Z_1 -1 \\
 \delta_2 &\equiv Z_2 -1 \\
 \delta_3 &\equiv Z_3 -1 \\
 \delta m &\equiv Z_2 m_0-m\ \
\end{align}

Thus we obtain the renormalized lagrangian at 1-loop with the appropriate counterterms:

\begin{align}
\mathcal{L}^{(1)} = \mathcal{L}+\Delta \mathcal{L}^{(1)}
= -\frac{1}{4} F_{\mu \nu} F^{\mu \nu}+\bar{\psi}(i \slashed{\partial}-m) \psi-e \bar{\psi} \gamma^\mu \psi A_\mu 
 -\frac{1}{4} \delta_3 F_{\mu \nu} F^{\mu \nu}+\bar{\psi}\left(i \delta_2 \slashed{\partial}-\delta m\right) \psi-e \delta_1 \bar{\psi} \gamma^\mu \psi A_\mu.
\label{eq:QEDL1loop}\end{align}
Eq. \ref{eq:QEDL1loop} introduces new Feynman rules related to the counterterms, as shown in Fig. \ref{fig:qedcounterterms}, responsible for the cancellation of the 1-loop divergences at all orders.

\begin{figure}
\begin{align}
& \begin{tikzpicture}
\begin{feynman} [small]
\node at (-3,0) {\bf $1.$};
\node (a) at (-2,0) ;
\node [crossed dot] (b) at (0,0);
\node  (c) at (2,0);
\node[right] (d) at (5,0) { $=-i \,(g_{\mu\nu} q^2-q_\mu q_\nu) \, \delta_3 $};
\diagram*{(a)--[photon](b)--[photon](c)} ;
\end{feynman} 
\end{tikzpicture} & \nonumber \\
& \begin{tikzpicture}
\begin{feynman}[small]
\node at (-3,0) {\bf $2.$};
\node (e) at (-2,0) ;
\node [crossed dot] (f) at (0,0);
\node (h) at (2,0);
\node[right] (w) at (5,0) { $= i \, (\delta_2 \slashed{p}-\delta m)$};
\diagram*{(e)--[fermion](f)--[fermion](h)} ;
\end{feynman} 
\end{tikzpicture} &   \nonumber  \\ 
& \begin{tikzpicture}
\begin{feynman}[small]
\node at (-3,0) {\bf $3.$};
\node (g) at (-2, 0);
 \node[crossed dot] (z) at (0, 0);
\node (r) at (-2,1);
\node (t) at (-2,-1);
\node (s) at (2,0) ;
\node[right] (y) at (5,0) {$=-i \, e \gamma^\mu \delta_1 $};
\diagram* {
(r) -- [fermion]  (z)  ;
(t)--[anti fermion] (z) ;
(z) -- [photon] (s);
};  
\end{feynman}
\end{tikzpicture} & \nonumber
\end{align}
	\caption{ Feynman diagrams for the counterterms in the renormalized QED lagrangian for: 1. the photon propagator, 2. the fermion propagator and 3. the vertex,  respectively.}
\label{fig:qedcounterterms}
\end{figure}
The values of the counterterms are set by imposing the renormalisation conditions:
\begin{align}
 \quad \Sigma(\slashed{p}= m)   & = 0,\\
 \quad \left.\frac{\Sigma(\slashed{p})}{d \slashed{p}}\right|_{\slashed{p}=m}  & =0,\\
 \quad\quad  \Pi(q^2=0) &=0,\\
 \quad -ieZ_1\Gamma^\mu (p-p^\prime=0)  &=-ie \gamma^\mu \label{eq:vertexren}
\end{align}
that fix the values of the renormalized mass, $m$, the renormalisation constants for the fermion and the photon field, the electron charge $e$ set at the scale $q^2 = 0$.
These are applied to the 1PI diagrams in Fig. \ref{fig:1piqed}, as shown in the previous section.

\begin{figure}
\begin{align}
& \begin{tikzpicture}
\begin{feynman} [small]
\node at (-3,0) {\bf $1.$};
\node (a) at (-2,0) ;
\node at (-1.5,0.2) {\contour{white}{\scriptsize \bf $\mu$}};
\node at (-1,-0.2) {\contour{white}{\scriptsize \bf $q$}};
\node [blob] (b) at (0,0);
\node at (0,0) {\contour{white}{\scriptsize \bf 1PI}};
\node  (c) at (2,0);
\node at (1.5,0.2) {\contour{white}{\scriptsize \bf $\nu$}};
\node at (1.0,-0.2) {\contour{white}{\scriptsize \bf $q$}};
\node[right] (d) at (5,0) { $= i \,(g_{\mu\nu} q^2-q_\mu q_\nu) \, \Pi(q^2) $};
\diagram*{(a)--[photon](b)--[photon](c)} ;
\end{feynman} 
\end{tikzpicture} & \nonumber \\
& \begin{tikzpicture}
\begin{feynman}[small]
\node at (-3,0) {\bf $2.$};
\node (e) at (-2,0) ;
\node at (-1,-0.2) {\contour{white}{\scriptsize \bf $p$}};
\node [blob] (f) at (0,0) {\contour{white}{\scriptsize \bf 1PI}};;
\node (h) at (2,0);
\node at (1,-0.2) {\contour{white}{\scriptsize \bf $p$}};
\node[right] (w) at (5,0) { $= - i \, (\Sigma( \slashed{p})$};
\diagram*{(e)--[fermion ](f)--[fermion](h)} ;
\end{feynman} 
\end{tikzpicture} & \nonumber \\
& \begin{tikzpicture}
\begin{feynman}[small]
\node at (-3,0) {\bf $3.$};
\node (g) at (-2, 0);
 \node[blob] (z) at (0, 0) {\contour{white}{\scriptsize \bf 1PI}};;
\node (r) at (-2,1);
\node at (-1.3,0.9) {\contour{white}{\scriptsize \bf $p$}};
\node (t) at (-2,-1);
\node at (-1.3,-0.9) {\contour{white}{\scriptsize \bf $p^\prime$}};
\node (s) at (2,0) ;
\node at (1,-0.2) {\contour{white}{\scriptsize \bf $q=p-p^\prime$}};
\node[right] (y) at (5,0) {$=-i \, e \Gamma^\mu (p,p^\prime)$};
\diagram* {
(r) -- [fermion]  (z)  ;
(t)--[anti fermion] (z) ;
(z) -- [photon] (s);
};  
\end{feynman}
\end{tikzpicture}  & \nonumber
\end{align}
\caption{ Feynman diagrams for the 1PI Green’s functions: 1. the photon propagator, 2. the fermion propagator and 3. the vertex, respectively.}
\label{fig:1piqed}
\end{figure}

BPHZ renormalisation is applied directly to the integrand and thus it doesn’t require explictily regularisation and is independent of the UV regulator. Morever  given that the subtractions are made at zero momentum,  the BPHZ-prescription is entangled with IR-behavior of a theory, thus it becomes difficult to extend to the case of a massless theory or of a theory that has a singular IR-behavior. Besides, in gauge-theories one has to preserve the original symmetries of the initial Lagrangian, because of the conservation laws, thus the BPHZ prescription becomes quite complex, because counterterms have to satisfy symmetry-requirements at all orders. For a more exhaustive discussion on the BPHZ renormalisation see Ref. \cite{Collins:1984xc}.

\subsection{Renormalisation and symmetries}

Renormalisation introduces new operators via the insertion of the counterterms, together with the scale and scheme dependence in the Lagrangian, and this may affect the conserved quantities of the original bare-Lagrangian. For example, considering the case of a massless theory, the bare-Lagrangian is in fact completely {\it conformal}, but after renormalisation a scale $\mu^2$-dependence is introduced, this is known as {\it dimensional transmutation}\cite{Coleman:1973sx}. Conserved quantities, such as charges $Q^i$, are related to conserved currents $J^i_\mu(x)$ derived by a symmetry or either way by an invariance of the Lagrangian under group transformations (Noether’s theorem \cite{Noether:1918zz}). For instance QED is invariant under local $U(1)$-gauge transformations, but it is also invariant under $U(1)$-global transformations. The latter are responsible for charge conservation. In general the Lagrangian in QFT theories are invariant under global space-time transformations (Poincar\'e group) and the gauge group, i.e.  Lie-groups ($SU(N)$ or $U(1)$), other symmetries might also occur such as global chiral symmetries, and supersymmetry which includes also the BRS-symmetry (Becchi, Rouet and Stora \cite{Becchi:1974md}). This subject is treated in more detail in the
chapter on {\it Symmetries and conservation laws}.
Hence, in the presence of symmetries, and therefore of conserved currents, not all counterterms are independent. 
Original symmetries of the Lagrangian manifest themselves also perturbatively by relations among correlation functions. These are known in QED as Ward-Takahashi identities\cite{Ward:1950xp,Takahashi:1957xn}, they are preserved perturbatively at all orders and they are not spoilt by renormalisation. These identities originate relations among the counterterms.

%%%%%

The QED Lagrangian of Eq. \ref{eq:Lqed}, is invariant under the global $U(1)$-group. The conserved vector current results:

\begin{align}
& J^\mu(x)  =e \bar{\psi}(x) \gamma^\mu \psi(x),\\
 &\partial_\mu J^\mu(x)  =0 
\label{eq:qedcc}
\end{align}
which implies the conservation of the charge $Q$ via the Gauss divergence theorem:

\begin{align}
\frac{dQ}{dt}  =\int_V d^3x \frac{\partial j^0(x)}{\partial t}=-\int_V d^3 \, x  \nabla \cdot {\bf j} =-\int_{\partial V} d^2 {\rm \bf s} \cdot {\bf j}.
\label{eq:qcons}
\end{align}
where the surface integral in the last term is expected to be null as the fields tend to zero sufficiently fast at infinity.
The Ward identity is given by the relation among the correlation function obtained in the vector current insertion, as shown in the three-point Green’s function in Fig.\ref{fig:wardid}:

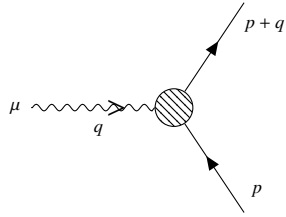
\begin{figure}[h]  
\begin{align}
& \begin{tikzpicture}
\begin{feynman}[small]
\node (g) at (0, 0);
\node (r)[blob] at (2,0);
\node (s) at (3,1.5);
\node (t) at (3,-1.5);
\node at (3.2,1.1) {\contour{white}{\scriptsize \bf $p+q$}};
\node at (3.1,-1.1) {\contour{white}{\scriptsize \bf $p$}};
\node at (1,-0.3) {\contour{white}{\scriptsize \bf $q$}};
\node at (1.2,0) {\contour{white}{  \Large  $>$}};
\node at (-0.1,0) {\contour{white}{\scriptsize \bf $\mu$}};
\diagram*{
(r) -- [fermion]  (s)  ;
(r)--[anti fermion] (t) ;
(g) -- [photon] (r);
};  
\end{feynman}
\end{tikzpicture} \nonumber \\
\end{align}
\caption{Feynman diagram of two fermion fields coupled to a vector current. \hspace{8cm}}	
\label{fig:wardid}
\end{figure}

\begin{align}
-iq^\mu G_\mu (p, q) = \Delta (p ) - \Delta (p+q), 
\end{align}
this relation holds also for the renormalized correlation functions:
\begin{align}
-iq^\mu G^R_\mu (p, q) = \Delta_R (p ) - \Delta_R (p+q), 
\end{align}
which underlies that the conserved currents do not renormalize as composite operators: i.e. $Z_J=1$
\footnote{This is a subtle passage in QED, in fact there might be the case of operators such as $\partial_\mu F^{\mu \nu}$, that disappear in the integration, but they may modify the conserved current non-renormalizability, however these terms do not spoil the Ward identity, see e.g. Ref. \cite{Collins:2005nj}}.
 By using Eq. \ref{eq:gammar} we can determine the Ward-identity for the 1PI Green’s function:
\begin{align}
-i q^\mu \Gamma^R_\mu(p, q)= \Delta_R^{-1}(p+q)- \Delta_R^{-1}(p),
\label{eq:gammar2}
\end{align}
where $ \Delta_R(p)$ is the renormalized fermion propagator in momentum space.
This Ward identity determines the relation between the counterterms $\delta_1$ and $\delta_2$ associated with the vertex and fermion propagator. In fact, from Eq. \ref{eq:gammar2} substituting the vertex renormalisation Eq. \ref{eq:vertexren} and using the renormalized fermion propagator:
\begin{align}
\Delta_R(p)=\frac{i Z_2}{\slashed{ p}-m+i \epsilon},
\end{align}
we obtain:

\begin{align}
Z_1 = Z_2 ,
\end{align}

Thus, the vector Ward identity implies that the renormalisation of the charge and of the fermion field originate the same counterterms:
\begin{align}
\delta_1=\delta_2.
\end{align}

It follows from Eq. \ref{eq:qedcoupling} that the renormalized charge:
\begin{align}
e= e_0 Z^{1/2}_3,
\end{align}
which implies that the charge renormalisation is independent of the fermion field and vertex renormalisations and it is entirely determined by the
renormalisation of the photon field only. 

Besides, the Ward identity is responsible for the convergence or for either a milder divergence of some graphs that by power counting should diverge with a larger superficial degree of divergence $D_s$. For example the light-by-light scattering has $D_s=0$, but is finite in QED, while the vacuum polarization diagram diverges only logarithmically, though has a $D_s=2$.
Thus, the original symmetries of the Lagrangian must be preserved also from the renormalized Lagrangian and this implies that operators introduced with counterterms have the same form and dimensions of the operators entering the bare Lagrangian. Besides, in gauge theories with {\it spontaneous symmetry breaking} (SSB), the SSB mechanism doesn’t spoil the renormalizability as shown by ’t Hooft and Veltman\cite{tHooft:1972tcz}. A different situation occurs for the case of the {\it effective field theories} (EFT) (e.g. see Ref. \cite{Manohar:2018aog}) or for the {\it operator product expansion} (OPE) , where operators with higher dimensions are introduced and mix under renormalisation \cite{Peskin:1995ev}.
%%%%%%%%%%%%%%%%%%%%%%%%

On the contrary, the Axial-vector-current Ward identity is not preserved perturbatively not even for a chiral Lagrangian, due to the occurrence of the ABJ-anomaly (Adler, Bell and Jackiw \cite{Adler:1969gk} \cite{Bell:1969ts}) from the triangular diagrams:

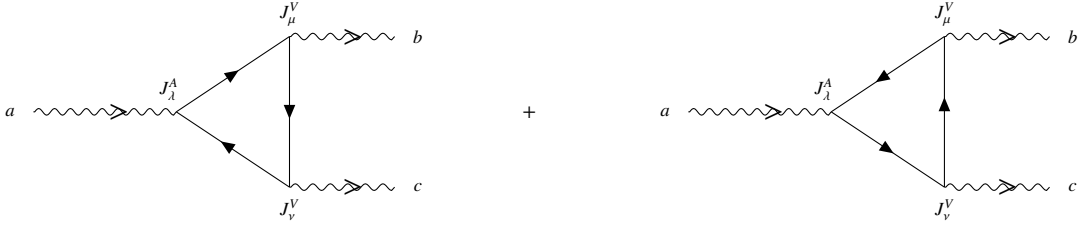
\begin{figure}[h]  
\begin{align}
& \begin{tikzpicture}
\begin{feynman}[small]
\node (a) at (0, 0);
\vertex (b) at (2,0);
\vertex  (c) at (3.5,1);
\vertex  (d) at (3.5,-1);
\node (e) at (5,1);
\node (f) at (5,-1);
\node  at (5.2,1) {\scriptsize \bf $b$};
\node  at (5.2,-1) {\scriptsize \bf $c$};
\node at (-0.2,0) {\scriptsize \bf $a$};
\node at (1.9,0.3) {\scriptsize \bf $J^A_\lambda$};
\node at (3.5,1.3) {\scriptsize \bf $J^V_\mu$};
\node at (3.5,-1.3) {\scriptsize \bf $J^V_\nu$};
\node at (1.2,0) {\contour{white}{  \Large  $>$}};
\node at (4.3,1) {\contour{white}{  \Large  $>$}};
\node at (4.3,-1) {\contour{white}{  \Large  $>$}};
\diagram*{
(a) -- [photon]  (b)  ;
(b)--[anti fermion] (d) ;
(b)--[fermion] (c) ;
(c)--[fermion] (d) ;
(d) -- [photon] (f);
(c) -- [photon] (e);
};  
\end{feynman}
\end{tikzpicture} \nonumber 
 & \begin{tikzpicture}
\begin{feynman}[small]
\node (a) at (-2, 0){+};
\node (a) at (0, 0);
\vertex (b) at (2,0);
\vertex  (c) at (3.5,1);
\vertex  (d) at (3.5,-1);
\node (e) at (5,1);
\node (f) at (5,-1);
\node  at (5.2,1) {\scriptsize \bf $b$};
\node  at (5.2,-1) {\scriptsize \bf $c$};
\node at (-0.2,0) {\scriptsize \bf $a$};
\node at (1.9,0.3) {\scriptsize \bf $J^A_\lambda$};
\node at (3.5,1.3) {\scriptsize \bf $J^V_\mu$};
\node at (3.5,-1.3) {\scriptsize \bf $J^V_\nu$};
\node at (1.2,0) {\contour{white}{  \Large  $>$}};
\node at (4.3,1) {\contour{white}{  \Large  $>$}};
\node at (4.3,-1) {\contour{white}{  \Large  $>$}};
\diagram*{
(a) -- [photon]  (b)  ;
(b)--[ fermion] (d) ;
(b)--[anti fermion] (c) ;
(c)--[anti fermion] (d) ;
(d) -- [photon] (f);
(c) -- [photon] (e);
};  
\end{feynman}
\end{tikzpicture} \nonumber 
\end{align}
\caption{Feynman diagrams responsible for the anomaly of a gauge symmetry current in a chiral gauge theory. \hspace{8cm}}	
\label{fig:anomaly}
\end{figure}

The contributions of the two diagrams in Fig. \ref{fig:anomaly}, does not cancel and in the Abelian case they result:
\begin{align}
\partial^\lambda J^A_\lambda=\frac{1}{16 \pi^2} \varepsilon^{\alpha\beta\gamma\delta}F_{\alpha\beta}F_{\gamma \delta},
\label{eq:axialward}
\end{align}
with $J^A_\mu=\bar{\psi} \gamma_\mu \gamma_5 \psi$ and $J^V_\mu=\bar{\psi} \gamma_\mu  \psi$ the axial and vector currents respectively.

These terms spoil the renormalizability and thus must cancel\cite{Gross:1972pv}\cite{Georgi:1972bb}. In QCD the anomaly is removed since the gauge coupling in QCD does not involve axial currents, or either way left-handed and right-handed quark currents equally couple cancelling out their opposite contributions to the QCD anomaly. On the other hand, in the electroweak $SU(2)_L \times U_Y$ theory, anomalies exactly cancel due to the particular structure of the families and to their quantum numbers including {\it color}. For a non-Abelian gauge theory the anomalous term Eq. \ref{eq:axialward} has an extra factor proportional to the trace ${\rm tr} (T^a(R)\{T^b(R), T^c(R)\})$, where the $T^a(R)$ is the group generator in the $R$ fermion representation. Thus, it follows that for any fermion representation $R$ the trace ${\rm tr}(T^a(R)\{T^b (R), T^c(R)\})$ must vanish, or at least the sum over all possible fermionic states (which includes left- and right- leptons and quarks), has to cancel. 
Thus, the electroweak $SU(2)_L \times U_Y$-sector of the Standard Model is free of anomalies and so is the entire SM (e.g. see Ref.\cite{Ellis:1990pp}).

%%%%%%%%%

%%%%%%%

\section{Renormalisation in QCD}

Once renormalisation has been introduced, the QCD Lagrangian can be written in the form~\cite{Chetyrkin:2004mf}:
\begin{align}
\mathcal{L} & =Z_2 \sum_{f=1}^{n_f} \bar{\psi}^f\left(\mathrm{i}  \slashed{\partial}+g Z_1^{\rm qqg} Z_2^{-1} \slashed{A}-Z_m m_f\right) \psi^f-\frac{1}{4} Z_3\left(\partial_\mu A_\nu-\partial_\nu A_\mu\right)^2-\frac{1}{2} g Z_1^{\rm 3 g}\left(\partial_\mu A_\nu^a-\partial_\nu A_\mu^a\right)\left(A_\mu \times A_\nu\right)^a \nonumber \\
& \quad -\frac{1}{4} g^2 Z_1^{\rm 4 g}\left(A_\mu \times A_\nu\right)^2- \frac{Z_3}{ Z_{\xi}} \frac{1}{2 \xi_L}\left(\partial_\nu A_\mu\right)^2+Z_3^{\rm c} \partial_\nu \bar{c}\left(\partial_\nu c\right) 
+g Z_1^{\rm c c g} \partial^\mu \bar{c}(A \times c)
\label{eq:qcdL}
\end{align}
where $m_f$ is the $f$-quark mass, $\psi_i^f$ is the $f$-quark field given in the fundamental $\mathrm{SU}(3)$ representation, $A_\mu^a$ is the gluon field given in the adjoint $\mathrm{SU}(3)$ representation.  The $c^a$ are the ghost fields and $\xi_L$ is the gauge parameter ( where $\xi_L=0$ corresponds to the Landau gauge).
The $Z_i$ are the RC constants determined by renormalisation counterterms for the fields and vertices.

In particular, $Z_3, Z_2, Z_3^{\rm c}$ are the renormalisation constants relating bare and renormalized fields :

\begin{align*}
A_0^{a \mu}=Z_3^{1 / 2} A^{a \mu}, \quad \psi_0^f=Z_2^{1 / 2} \psi^f, \quad c_0^a=(Z_3^{\rm c})^{1 / 2} c^a,
\end{align*}
the gluon, quark and ghosts fields, respectively, while:
\begin{align*}
Z_1^V, \quad V \in\{\mathrm{3g},  \mathrm{4g}, {\rm ccg }, \mathrm{qqg}\}
\end{align*}

are the RCs for the renormalisation of the 3-gluon, 4-gluon, ghost-ghost-gluon, quark-quark-gluon vertex respectively. $Z_{\xi}$ is the RC for the gauge fixing parameter $\xi_L$. 
For QCD, these quantities are often given in the {\it  minimal subtraction scheme } ($\rm MS$)\cite{tHooft:1973mfk,Weinberg:1973xwm}, i.e. the scheme defined by the only subtraction of the pole occurring in dimensional regularisation,
$1/\varepsilon$. A
more suitable scheme is provided by the {\it modified minimal subtraction }$\overline{\rm
MS}$~\cite{Bardeen:1978yd}, where also the constant term $\ln(4 \pi)-\gamma_E$ is subtracted out together with the pole. These schemes differ from the momentum subtraction scheme (MOM) that we have introduced in the previous section. However, at the {\it next-to-leading-order} (NLO) of accuracy, different schemes can be related by a scale redefinition, i.e. by scale transformation, e.g.
$\mu^2 \rightarrow 4 \pi \mu^2 e^{-\gamma_E}$, by using the renormalisation Group. 
In general, different schemes can be related at all orders, by using the {\it extended renormalisation group} transformations, that will be introduced in the following section. 
The physical quantities in the QCD Lagrangian are thus defined at a given subtraction point, i.e. the renormalisation scale $\mu$, and using a particular scheme, e.g. $\overline{\rm
MS}$. It follows that a scale-and-scheme dependence is introduced in the theory and this characterises the particular value of the physical parameters, such as the renormalized coupling $\alpha_s^{\overline{\rm MS}}(\mu)$:

\begin{align}
\alpha_s^{\overline{\rm MS}}\left(Q \right)=Q^{-2 \varepsilon} Z^{-1}_{\alpha_s}\left(Q \right) {\alpha_s^0}
\label{eqn:alphasren} 
 \end{align}
where $\alpha_s\equiv \frac{g^2}{4\pi}$.
We remark that the renormalisation procedure
leads to a unique renormalisation constant $Z_{\alpha_s}\equiv
Z^2_g $ for the strong coupling. In fact, the other
renormalisation constants, such as $Z_1^V$, $V\in({\rm 3g, 4g, ccg,
qqg})$ are related to the coupling’s RC, via the Slavnov--Taylor identities
\cite{Slavnov:1972fg} \cite{Taylor:1971ff} (for a review see e.g. \cite{Chetyrkin:2004mf}). 
These relations are given by:
\begin{align}
Z_{\xi}& = Z_3,  \label{eq:STid1}\\
Z_g & =  Z_1^{\rm 3g}\left(Z_3\right)^{-3/2},  \\
Z_g & =  \sqrt{Z_1^{\rm 4g}}\left(Z_3\right)^{-1},   \\
Z_g & =  Z_1^{\rm ccg}\left(Z_3\right)^{-1/2}\left(Z_3^{\rm c}\right)^{-1}, \\
Z_g & =  Z_1^{\rm qqg}\left(Z_3\right)^{-1/2}\left(Z_2\right)^{-1}.
\label{eq:STid}
\end{align}

Thus, it is possible to express all vertex RCs in terms of $Z_{\alpha_s}$, and of the fields RCs. This is an effect of the renormalizability of the QCD, which avoids the occurrence of further singularities at higher orders that cannot be reabsorbed into the initial set of renormalized parameters, but would need new further parameters to be introduced in order to be cancelled.

\subsection{The renormalisation Group \label{RGroup}}

We show in this section how the RG equations for the coupling can be derived in QCD.
We refer to the case of adopting the dimensional regularisation procedure and the minimal subtraction scheme. Thus, we introduce the dimensional regulator $\varepsilon$, $D=4-2\varepsilon$ and we apply renormalisation adopting the ${\rm MS}$-scheme at the scale $\mu=Q$. 
 Considering the Slavnov-Taylor identity in Eq. \ref{eq:STid}:

\begin{align}
Z^{-1}_{\alpha_s}=(\sqrt{Z_3} Z_2 /Z_1)^2,
 \label{eqn:zalphas} 
 \end{align}
where we have dropped the superscript in the  ${\rm qqg}$-vertex RC, i.e. $Z_1\equiv Z_1^{\rm qqg}$, and considering the
 renormalisation constants: \begin{align} Z_1(Q)&=
1-\frac{\alpha_s(Q)}{4\pi} \big(N_c+C_F\big) \frac{1}{\varepsilon} \\
Z_2(Q)&= 1-\frac{\alpha_s(Q)}{4\pi} C_F  \frac{1}{\varepsilon} \\
Z_3(Q) &= 1+\frac{\alpha_s(Q)}{4\pi}
\left(\frac{5}{3}N_c-\frac{2}{3}N_f\right) \frac{1}{\varepsilon}
  \end{align}
where $C_F=\frac{(N_c^2-1)}{2N_c}$ and $N_c, N_f$ are numbers of colors and the number of active quark flavors at the physical scale $Q$, we obtain: 

\begin{align}
Z_{\alpha}\left(Q \right)=1-\frac{\alpha_{s}\left(Q \right)}{4
\pi} \beta_{0} \frac{1}{\varepsilon},  \end{align} with \begin{align}
\beta_{0}=11-\frac{2 N_f}{3}, \label{eq:zetabeta0}  \end{align}
where $\beta_0$ is the first coefficient of the $\beta$-function in Eq. \ref{eqn:betag}.

From the relation between the renormalized strong coupling $\alpha_s(Q)$ and the bare
coupling $\alpha^0_{s}$, Eq. \ref{eqn:alphasren}, it follows that the values of the coupling at two different scales can be related by:
 \begin{align}
{\alpha^0_{s}}=Q^{2 \varepsilon} Z_{\alpha}\left(Q \right)
\alpha_{s}\left(Q \right)=\mu^{2 \varepsilon} Z_{\alpha}\left(\mu
\right) \alpha_{s}\left(\mu \right),
 \label{eqn:alphascales}  \end{align}

The transformations of the RG for the coupling can be derived from Eq.~\ref{eqn:alphascales}, and thus we can obtain the relation from two different couplings at two
different scales: 
\begin{align}
\alpha_{s}\left(Q\right)=\mathcal{Z}_{\alpha}\left(Q,
\mu\right) \alpha_{s}\left(\mu\right), \end{align} 
with
\begin{align}
\mathcal{Z}_{\alpha}\left(Q, \mu\right)
\equiv\left(\mu^{2 \varepsilon} / Q^{2
\varepsilon}\right)\left[Z_{\alpha}\left(\mu \right) /
Z_{\alpha}\left(Q\right)\right].
\end{align}

The $\mathcal{Z}_{\alpha}$ form a group with a composition law:
\begin{align}\mathcal{Z}_{\alpha}\left(Q,
\mu\right)=\mathcal{Z}_{\alpha}\left(Q, \mu_{0}\right)
\mathcal{Z}_{\alpha}\left(\mu_{0}, \mu\right),\end{align} a unity
element: $\mathcal{Z}_{\alpha}\left(Q, Q\right)=1$ and an
inversion law: $\mathcal{Z}_{\alpha}\left(Q,
\mu\right)=\mathcal{Z}_{\alpha}^{-1}\left(\mu,
Q\right)$. Analogous RG transformations are determined for the other independent renormalized parameters: masses and fields (see e.g. Refs.  \cite{Collins:1984xc,Cheng:1984vwu}). Fundamental properties of the renormalisation group
are: {\it reflexivity, symmetry } and {\it transitivity}.

\subsubsection{The QCD $\beta(\alpha_s)$-function }

 We discuss in this section the dependence of the
renormalized coupling $\alpha_s(Q^2)$ on the scale $Q^2$. As shown
in QED by Gell-Mann and Low\cite{Gell-Mann:1954yli}, this dependence can be described
introducing the $\beta$-function:
 \begin{align}
\frac{1}{4 \pi}\frac{d\alpha_s(\mu)}{d \log \mu^2}=\beta(\alpha_s(\mu)),
\label{betafun1}\end{align} where $\alpha_s(\mu)\equiv \frac{g^2(\mu)}{4\pi}$ and \begin{align}
\beta\left(\alpha_{s}\right)=-\left(\frac{\alpha_{s}}{4
\pi}\right)^{2} \sum_{n=0} \left(\frac{\alpha_{s}}{4
\pi}\right)^{n} \beta_{n}. \label{betafun10}\end{align}

Neglecting quark masses, the first two $\beta$-terms are RS
independent and they have been calculated in
Refs.~\cite{Gross:1973id,Politzer:1973fx,Caswell:1974gg,Jones:1974mm,Egorian:1978zx}
for the $\overline{\rm MS}$ scheme:
\begin{align}\beta_{0}=\!\frac{11}{3}C_{A}\!-\!\frac{4}{3}T_{R}N_{f},\end{align}
\begin{align}\beta_{1}=\!
\frac{34}{3}C_{A}^{2}\!-\!4\left(\frac{5}{3}C_{A}\!+\!C_{F}\right)T_{R}N_{f} \end{align}
where $C_F=\frac{\left(N_{c}^{2}-1\right)}{2 N_{c}}$, $C_A=N_c$
and $T_R=1/2$ are the color factors for the ${\rm SU(3)}$ gauge group~\cite{Mojaza:2010cm}. \\
At higher loops we have that $\beta_2,\beta_3,\beta_4,\dots$ are scheme dependent and results for $\overline{\rm MS}$ have been calculated up to five-loop of accuracy:
\begin{align}
\beta_{2}&\simeq1428.5 - 279.6\, N_f + 6.0\, N_f^2 ,\\
\beta_{3}&\simeq 29242.9 - 6946.3 \, N_f + 405.1 \, N_f^2 + 1.5 \,N_f^3 ,\\
\beta_{4}&\simeq 537149.4 - 186163.2 \, N_f + 17571.8 \, N_f^2 - 231.3 \, N_f^3 - 1.8 \, N_f^4,
\label{eq:beta234}   
\end{align}
in Refs.~\cite{Larin:1993tp},\cite{vanRitbergen:1997va},\cite{Baikov:2016tgj} respectively.

Thus, the strong coupling RC (Eq. \ref{eq:zetabeta0}) becomes: 
 \begin{align}
Z_{a}(\mu)&= 1-\frac{\beta_{0}}{\epsilon} a+\left(\frac{\beta_{0}^{2}}{\epsilon^{2}}-\frac{\beta_{1}}{2 \epsilon}\right) a^{2}-\left(\frac{\beta_{0}^{3}}{\epsilon^{3}}-\frac{7}{6} \frac{\beta_{0} \beta_{1}}{\epsilon^{2}}+\frac{\beta_{2}}{3 \epsilon}\right) a^{3}  +\left(\frac{\beta_{0}^{4}}{\epsilon^{4}}-\frac{23 \beta_{1}
\beta_{0}^{2}}{12 \epsilon^{3}}+\frac{5 \beta_{2} \beta_{0}}{6
\epsilon^{2}}+\frac{3 \beta_{1}^{2}}{8
\epsilon^{2}}-\frac{\beta_{3}}{4 \epsilon}\right) a^{4}+\cdots,
\label{eqn:zexp}
\end{align}
where $a=\alpha_s(\mu)/(4 \pi)$. 
In QCD the number of colors $N_c$ is set to 3 by the gauge group, while $N_f$, i.e. the
number of active flavors, varies with the scale $Q$ across quark
thresholds. Thus, the values of the $\beta_i$ coefficients vary with the number of active flavors $(0\leq N_f \leq 6)$ entering the loop integral at a given energy scale.
Given the scale dependence of the strong coupling, any experimental measurement is determined at a particular physical scale which is usually taken as the scale of the process:
$\alpha_s(Q)$.  In general the value which characterises the strong coupling phenomenologically, is the one determined at the scale $\alpha_s(M_Z)$, where $M_Z$ is the $Z^0$-mass scale (see e.g.  the Particle Data Group (PDG) \cite{ParticleDataGroup:2024cfk}).

\subsubsection{Analytical solution for $\alpha_s(\mu)$\label{oneloop}}

Analytical solutions for the truncated Eq.~\ref{betafun1} exist up to two-loop accuracy. 
At one-loop, the solution is given by : \begin{align}
\int_{\alpha_s(\mu_0^2)}^{\alpha_s(\mu^2)} \frac{1}{4\pi}
\frac{d\alpha_{s}}{\beta(\alpha_{s}) }=-\int_{\mu_0^2}^{\mu^2}
\frac{d Q^{2}}{Q^{2}}, \end{align}
which leads to:
 \begin{align} \frac{4
\pi}{\alpha_{s}\left(\mu_{0}^{2}\right)}-\frac{4
\pi}{\alpha_{s}\left(\mu^{2}\right)}=\beta_{0} \ln
\left(\frac{\mu_{0}^{2}}{\mu^{2}}\right). \label{1loopalphas} \end{align}
The solution is usually given in the more familiar explicit form:

 \begin{align}
\alpha_{s}(\mu^{2}) = \frac{\alpha_s(\mu^2_0)}{1+\beta_0 \frac{
\alpha_s(\mu_0^2)}{4 \pi} \ln(\mu^2/\mu_0^2)}. \label{1loopalphas2}
\end{align} This solution relates one known (measured value) of the
coupling at a given scale $\mu_0$ with an unknown value
$\alpha_s(\mu^2)$. 
Either way, the solution can be given
introducing the {\it QCD scale parameter} $\Lambda$, defined as: \begin{align}\Lambda^{2} \equiv \mu_0^{2}
e^{-\frac{4 \pi}{\beta_{0} \alpha_{s}\left(\mu_0^{2}\right)}}
\label{lambda1loop}\end{align} which yields the familiar one-loop
solution:
\begin{align}
\alpha_{s}\left(Q^{2}\right)=\frac{4 \pi}{\beta_{0} \ln
\left(Q^{2} / \Lambda^{2}\right)}.
\end{align}

Already at the one loop level one can distinguish two regimes of
the theory. For the number of flavors larger than $11N_c/2$ (i.e.
the zero of the $\beta_0$ coefficient) the theory possesses an
infrared non-interacting fixed point and at low energies the
theory is known as {\it non-abelian quantum electrodynamics}
(non-abelian QED). The high energy behavior of the theory is
uncertain, it depends on the number of active flavors and there is
the possibility that it could develop a critical number of flavors
above which the theory reaches an UV fixed
point~\cite{Antipin:2017ebo} and therefore becomes safe. When the
number of flavors is below $11 N_c /2$ the non-interacting fixed
point becomes UV in nature and then we say that the theory is {\it
asymptotically free}.

It is straightforward to check the asymptotic limit of the
coupling in the deep UV region: \begin{align}\lim_{s \rightarrow \infty}
\alpha_s(s)=0. \end{align} This result is known as {\it asymptotic
freedom} and it is the outstanding result that has justified QCD
as the most accredited candidate for the theory of strong
interactions. On the other hand, we have that the perturbative
coupling diverges at the $\Lambda \sim (200-300){\rm MeV}$ scale.
This is sometimes referred to as the {\it Landau ghost pole} to
indicate the presence of a singularity in the coupling that is
actually unphysical and indicates the breakdown of the
perturbative regime. However, by including non-perturbative contributions, or by using non-perturbative QCD, this singularity can be removed leading to the correct
finite limit at any $N_f$~\cite{Deur:2016tte}.

 The Landau-pole is not an explanation for
confinement, though it might indicate its presence. When the
coupling becomes too large the use of a non-perturbative approach to QCD is mandatory in order to obtain reliable results. We remark that the scale parameter $\Lambda$ is RS dependent and its definition depends on the order of accuracy of the coupling $\alpha_s(Q^2)$. 
Considering that the solution $\alpha_{s}$ at
order $\beta_{0}$ or $\beta_{1}$ is universal, the definition of
$\Lambda$ at the first two orders is usually preferred,

\begin{figure}[htb]
\centering
\includegraphics[width=8cm,height=6cm]{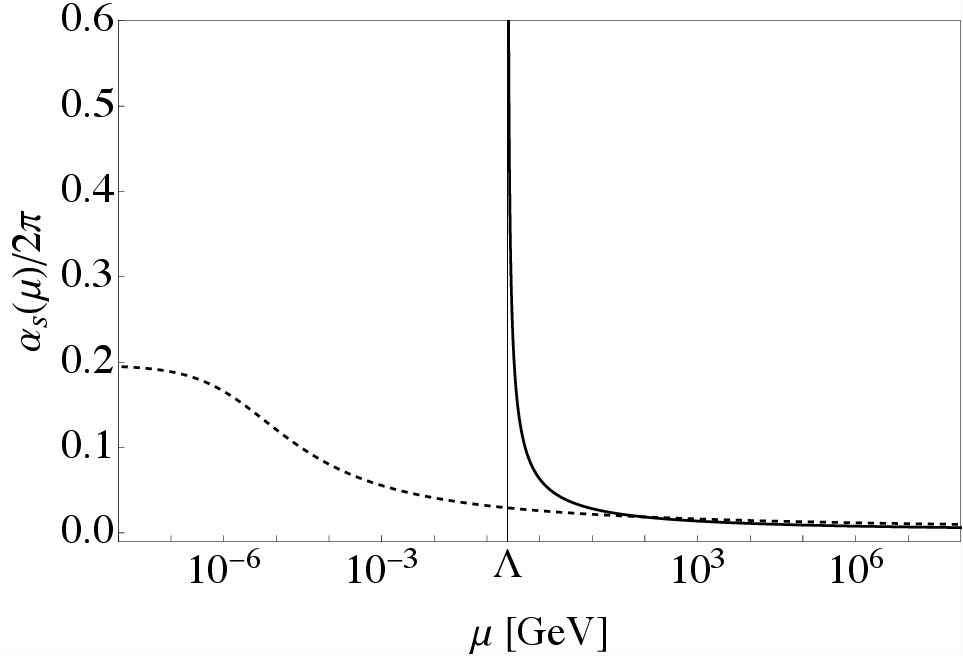}
\includegraphics[width=8cm,height=6cm]{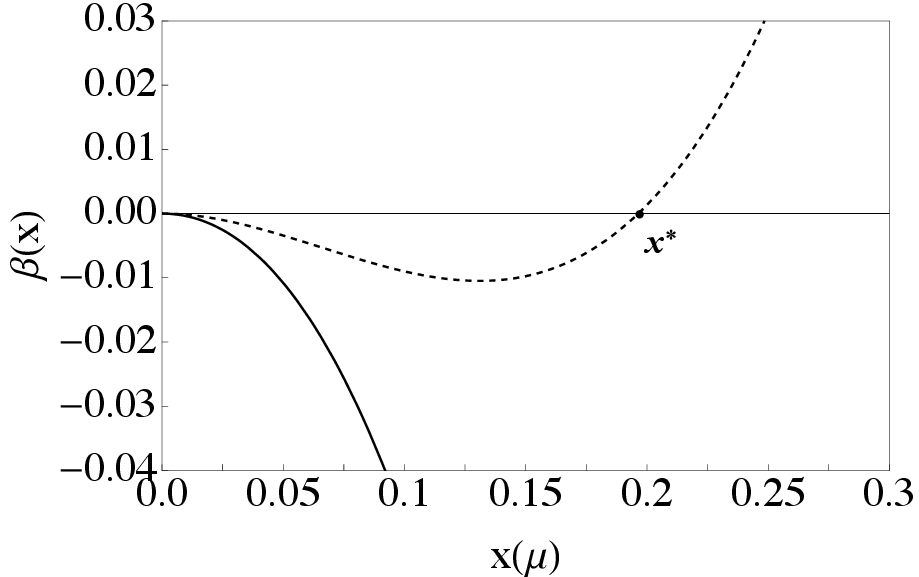}
\caption{The two-loop strong coupling $\alpha_s(\mu)$, on the left,  and the two-loop QCD $\beta$-function $\beta(x)$, with $x(\mu)\equiv \alpha_S(\mu)/(2\pi)$, for $N_f=11$
(black dashed) and for $N_f=5$ (solid
black). The interacting IR-fixed point $x^*$ is also shown.} \label{fig:lambert}
\end{figure}

\subsubsection{The conformal window of perturbative QCD\label{twoloops}}

%%%%%%%%%%%%%%%%%%%%%%%%%%%%%%%%%%%%
Including also the $\beta_1$ coefficient at two-loop accuracy, we still can solve the Eq.~\ref{betafun1} analytically.
In order to determine the solution for the strong coupling
$\alpha_s$ at the next-to-next leading order (NNLO), we introduce the following
notation: $x(\mu)\equiv \frac{\alpha_s(\mu)}{2 \pi}$,
$t=\log(\mu^2/\mu_0^2)$, $B=\frac{1}{2}\beta_0$ and
$C=\frac{1}{2}\frac{\beta_1}{\beta_0}$, $x^*\equiv -\frac{1}{C}$. By substituting these into Eq.~\ref{betafun1} , we obtain the differential equation:
\begin{equation}
\frac{dx}{dt}=-B x^2(1+C x) \label{lambert1}
\end{equation}
An implicit solution to Eq.~\ref{lambert1} is given by the Lambert
$W(z)$ function:
\begin{align}
W e^W = z \label{W}
\end{align}
with: $ W=\left(\frac{x^*}{ x}-1\right)$. The general solution for
Eq.~\ref{lambert1} is given by:
\begin{align}
x &= \frac{x^*}{1+W} , \label{eq:wlamb}\\
 z &= e^{\frac{x^*}{x_0}-1} \left(\frac{x^*}{x_0}-1 \right) \left( \frac{\mu^2}{\mu_0^2}
\right)^{x^* B}. \label{xz}
\end{align}
where
$x_0\equiv \alpha_s(M_Z) /(2\pi)= 0.01876 \pm 0.00016$ is
the coupling determined at the $Z^0$ mass
scale~\cite{ParticleDataGroup:2020ssz}.

%%%%%%%%%%%%%%%%%%%%%%%%%%
In this case, we have that not only the sign of $\beta_0$, but also the sign of $\beta_1$ characterises the solution.

depend on the values of the $N_c,\,N_f$. Given that the number $N_c$ is fixed by the group $\rm SU(N_c)$ of the theory, the only parameter that can vary is the number of active flavors $N_f$.

In fact, in the range i.e. $0 < N_f<\frac{34 N_c^3}{13 N_c^2-3}$, we notice that Eq.~\ref{lambert1} has a physical solution given by the  $W_{-1}$ branch. In this range the $\beta$-function is characterised by the non-interacting UV fixed point, as shown in Fig.~\ref{fig:lambert} (solid black curve on the right).
Thus, for physical values, i.e. $\left(0 \leq N_f \leq 6\right)<\frac{34 N_c^3}{13 N_c^2-3}$, we have that both $B>0, C>0$ are positive
and by introducing the standard QCD scale
parameter $\Lambda$ at two-loop:

\begin{align} \Lambda= \mu_0 \left(1+ \frac{|x^*|}{x_0}
\right)^{\frac{1}{2 B |x^*|}} e^{-\frac{1}{2 B x_0}}
\label{landaupole}\end{align}

we can write Eqs. \ref{eq:wlamb}-\ref{xz} as follows:

\begin{align}
x &= \frac{x^*}{1+W_{-1}} , \\
 z &= -\frac{1}{e}  \left( \frac{\mu^2}{\Lambda^2}
\right)^{x^* B}. \label{xz2}
\end{align}
This solution is shown in Fig.~\ref{fig:lambert} (solid black curve on the left), where it is also shown the Landau-pole at $\Lambda$.
On the other hand, as shown by Banks and Zaks in their analysis~\cite{Banks:1981nn}, when  $\frac{34 N_c^3}{13
N_c^2-3}< N_f<11/2 N_c$, the $\beta$-function develops both a non-interacting-UV and an interacting-IR fixed point: $x^*$, and the solution for the strong coupling is given by the $W_{0}$ branch in Eq.~\ref{eq:wlamb}.
The peculiar strong coupling IR-conformal behavior is shown in Fig.~\ref{fig:lambert} (dashed black curve on the left ), while the  $\beta$-function and its fixed points is shown in Fig.~\ref{fig:lambert} (dashed black curve on the right).
The two-dimensional region in the number of flavors and colors where asymptotically-free QCD develops an IR interacting fixed point is colloquially known as the {\it conformal window of pQCD}.
The two-loop solution for the strong coupling is discussed in more detail in Refs: \cite{Gardi:1998qr,DiGiustino:2021nep}.
In general IR and UV fixed points of the $\beta$-function can also be determined at different values of the number of colors $N_c$
(different gauge group $SU(N)$) and $N_f$ extending this analysis also to a wider set of gauge theories~\cite{Ryttov:2017khg}.

\subsection{Renormalisation group equations at high accuracy}

\subsubsection{The $\alpha_s(\mu)$ perturbative  solution}

At 3-loop, it is still possible to solve the truncated RG equation Eq.~\ref{betafun1} analytically, by introducing the  Pad\'e Approximant (PA)\cite{Basdevant:1972fe, Samuel:1992qg, Samuel:1995jc} for the $\beta$-function (this is shown in detail in Ref.~\cite{Gardi:1998qr}). However, the general perturbative solution for the strong coupling up to five-loop accuracy is obtained by 
%For orders up to $\beta_{4}$, an approximate analytical solution
%is obtained 
integrating Eq.~\ref{betafun1} :\begin{align}
 \ln \frac{\mu^{2}}{\Lambda^{2}} = \int \frac{d a}{\beta(a)}
\quad=\frac{1}{\beta_{0}}\left[\frac{1}{a}+b_{1} \ln a+a\left(-b_{1}^{2}+b_{2}\right)\right.
+a^{2}\left(\frac{b_{1}^{3}}{2}-b_{1} b_{2}+\frac{b_{3}}{2}\right)+a^{3}\left(-\frac{b_{1}^{4}}{3}+b_{1}^{2} b_{2}-\frac{b_{2}^{2}}{3}\left.\frac{2}{3} b_{1}
b_{3}+\frac{b_{4}}{3}\right)+O\left(a^{4}\right)\right]+C
\end{align}
where $a=\alpha_s(\mu)/(4 \pi)$, $C$ is a scheme factor, $b_N\equiv \beta_N/\beta_0$,
$(N=1,..,4)$ and performing the inversion of the last formula by
iteration as shown in Ref.~\cite{Kniehl:2006bg}, achieving the
result :
\begin{align}
a &= \frac{1}{\beta_{0} L}-\frac{b_{1} \ln L}{\left(\beta_{0} L\right)^{2}}+\frac{1}{\left(\beta_{0} L\right)^{3}}\left[b_{1}^{2}\left(\ln ^{2} L-\ln L-1\right)+b_{2}\right] +\frac{1}{\left(\beta_{0} L\right)^{4}}\left[b_{1}^{3}\left(-\ln ^{3} L+\frac{5}{2} \ln ^{2} L+2 \ln L-\frac{1}{2}\right)\right. \nonumber\\
&-\left. 3 b_{1} b_{2} \ln L+\frac{b_{3}}{2}\right]+\frac{1}{\left(\beta_{0} L\right)^{5}}\left[b _ { 1 } ^ { 4 } \left(\ln ^{4} L-\frac{13}{3} \ln ^{3} L\right.\right.-\left. \frac{3}{2} \ln ^{2} L+4 \ln L+\frac{7}{6}\right)+3 b_{1}^{2} b_{2}\left(2 \ln ^{2} L-\ln L-1\right) \nonumber \\
&-\left. b_{1} b_{3}\left(2 \ln L+\frac{1}{6}\right)+\frac{5}{3}
b_{2}^{2}+\frac{b_{4}}{3}\right]+O\left(\frac{1}{L^{6}}\right) .
\end{align}
where $L=\ln(\mu^2/\Lambda^2).$ The same definition of $\Lambda$
scale given in Eq.~\ref{landaupole} has been used for the
$\overline{\rm MS}$ scheme which leads to set the constant $C=
\left(b_{1} / \beta_{0}\right) \ln (\beta_{0})$.

%%%%%%%%%%%%%%%%%%%%%%%%%%%%%%%%%%%%

\subsubsection{The mass anomalous dimension}
The anomalous dimensions for mass and fields, that have been defined in Eq.~\ref{eq:anomalousphi} and Eq.~\ref{eq:anomalousm}, can be derived from their renormalisation constants $Z_m$, $Z_2,Z_3,Z_3^c$, that together with $Z_{\alpha_s}$ can be selected as the independent RCs. On the other hand, the other RCs can be derived using relations Eqs.~\ref{eq:STid1}-\ref{eq:STid}.
In general, the renormalisation constants $Z$s, do not depend on any dimensional parameter, such as mass or a particular momentum;
these can be written in the form of a double expansion in the couplant $a$ and in the reciprocal of the dimensional regulator $\varepsilon^{-1}$ as follows:
\begin{align}
Z\left(a,\frac{1}{\varepsilon}\right)=1+\sum_{i=1}^{\infty}\sum_{j=1}^{ i} Z_{i ,j} \frac{a^i}{\varepsilon^j}.
\label{eq:zij}
\end{align}

and considering that, keeping the previous normalisation, the $\beta$-function in $D=4-2\varepsilon$ dimensions results:
\begin{align}
\frac{d a}{d \log \mu^2}=-  \varepsilon \, a+\beta\left(a \right) \label{eq:betaeps}
\end{align}
with
\begin{align}
\beta\left(a\right)=- a \, \frac{d \log Z_\alpha\left(a\right)}{d \log \mu^2},
\end{align}

The RCs are related to the anomalous dimensions by:
\begin{align}
\gamma(a)=-\mu^2 \frac{\mathrm{~d} \log Z(a,\frac{1}{\varepsilon})}{\mathrm{d} \mu^2}=-\frac{\partial \log Z(a,\frac{1}{\varepsilon})}{\partial a}\frac{d a }{d \log \mu^2}=- \frac{\partial \log Z(a,\frac{1}{\varepsilon})}{\partial a}\left( -\varepsilon a+\beta(a) \right)=\sum_{n=1}^{\infty}Z_{n,1}\left(a,\frac{1}{\varepsilon}\right) \, n\, a^{n} . \label{eq:gammaRC}
\end{align}
Combining Eq. \ref{eq:gammaRC} and Eq. \ref{eq:betaeps}, we obtain the mass anomalous dimension:  
\begin{align}
 \gamma_m=-\sum_{n=0}^{\infty}\gamma_n a^{n+1}=\sum_{n=1}^{\infty}\left(Z_m \right)_{n,1}  \, n\, a^{n}\label{eq:gammam}
\end{align}

where the $Z_{n,1}$ are the coefficients related to the leading power $\frac{1}{\varepsilon}$-pole.
Up to five-loop accuracy the mass anomalous dimension is given by (see e.g Ref.~\cite{Baikov:2017ujl}):
\begin{align}
\gamma_m= & -4a-4^2\left(4.21-0.14 n_f\right)a^2  -4^3\left(19.52-2.28 n_f-0.03 n_f^2\right) a^3
-4^4\left(98.94-19.11 n_f+0.28 n_f^2+0.01 n_f^3\right)a^4 \nonumber \\
& -4^5\left(559.71-143.60 n_f+7.48 n_f^2+0.11 n_f^3-0.00008535 n_f^4\right) a^5+{\cal O}\, (a^6)\label{eq:gammanum}
\end{align}

%%%%%%%%%%%%%%%%%%%%%COPY-mass and Z at 5-loop
Analogously, it is straightforward to obtain the results for the anomalous dimensions $\gamma_2,\gamma_3,\gamma_3^c$ related to the fields RCs (see e.g. Refs.\cite{Chetyrkin:2017bjc,Luthe:2017ttc}).

%%%%%%%

\subsubsection{The running quark mass}

In order to obtain the perturbative solution for the evolution of the renormalised mass, we notice that dividing Eq. \ref{eq:anomalousm} by Eq. \ref{eq:betaeps} in the limit $\varepsilon\rightarrow 0$, we achieve:

\begin{align}
\frac{d \log m}{d  a}=\frac{\gamma_m\left(a\right)}{\beta\left(a\right)}. \label{eq:massa}
\end{align}

and integrating Eq. \ref{eq:massa} it is straightforward to obtain the solution for the mass renormalised at the scale $\mu$ :

\begin{align}
m\left(\mu \right)=m(\mu_0) \exp \int^{a(\mu)}_{a\left(\mu_0 \right)} \frac{\gamma_m\left(a\right)}{ \beta\left(a\right)}  d a \label{eq:mprime}
\end{align}

which can be written as:
\begin{align}
\frac{m(\mu)}{m\left(\mu_0\right)}=\frac{{\rm c} \left(a(\mu)\right)}{{\rm c}\left(a\left(\mu_0\right)\right)}, 
\end{align}
with
\begin{align}
{\rm c}(a)=\exp \left\{\int^a d a^{\prime} \frac{\gamma_m\left(a^{\prime}\right)}{\beta\left(a^{\prime}\right)}\right\},
\end{align}
One way to write the solution in Eq.~\ref{eq:mprime}, is obtained by using  the renormalisation group invariant (RGI) mass, which is defined as:
\begin{align}
\hat{m}_q = \frac{m_q\left(\mu_0\right)}{  {\rm c} \left(a \left(\mu_0\right)\right)},
\end{align}
this leads to the perturbative solution for the evolution equation~\cite{Vermaseren:1997fq} by expanding in powers of the coupling $a$:

\begin{align}
m_q\left(\mu\right)=\hat{m}_q \, {\rm c} (a)=\hat{m}_q\,  a^{\gamma_0 / \beta_0}\left[1+A_1 a+\left(A_1^2+A_2\right) \frac{a^2}{2}+\left(\frac{1}{2} A_1^3+\frac{3}{2} A_1 A_2+A_3\right) \frac{a^3}{3}+O\left(a^4\right)\right]
\end{align}

where the coefficients are:

\begin{align}
A_1 &=-\frac{\beta_1 \gamma_0}{\beta_0^2}+\frac{\gamma_1}{\beta_0} \\
A_2 &=\frac{\gamma_0}{\beta_0^2}\left(\frac{\beta_1^2}{\beta_0}-\beta_2\right)-\frac{\beta_1 \gamma_1}{\beta_0^2}+\frac{\gamma_2}{\beta_0} \\
A_3 & =\left[\frac{\beta_1 \beta_2}{\beta_0}-\frac{\beta_1}{\beta_0}\left(\frac{\beta_1^2}{\beta_0}-\beta_2\right)-\beta_3\right] \frac{\gamma_0}{\beta_0^2}+\frac{\gamma_1}{\beta_0^2}\left(\frac{\beta_1^2}{\beta_0}-\beta_2\right)-\frac{\beta_1 \gamma_2}{\beta_0^2}+\frac{\gamma_3}{\beta_0}
\end{align}

Thus, the running quark mass for the $b$ (bottom) and $t$ (top) quark in the $\overline{\mathrm{MS}}$ at four loop accuracy in terms of the invariant quark mass are given by:

\begin{align}
m_b\left(\mu\right)&\simeq\hat{m}_b\left(\frac{\alpha_s}{\pi}\right)^{12 / 23}\left[1+1.17549\left(\frac{\alpha_s}{\pi}\right)+1.50071\left(\frac{\alpha_s}{\pi}\right)^2+0.172478\left(\frac{\alpha_s}{\pi}\right)^3\right] \\
m_t\left(\mu\right)&\simeq\hat{m}_t\left(\frac{\alpha_s}{\pi}\right)^{4 / 7}\left[1+1.39796\left(\frac{\alpha_s}{\pi}\right)+1.79348\left(\frac{\alpha_s}{\pi}\right)^2-0.683433\left(\frac{\alpha_s}{\pi}\right)^3\right]
\end{align}
where different number of active flavors $N_f$ are introduced in the $\gamma_i$, $\beta_i$ coefficients, according to the value of the renormalisation scale. 
%m_s\left(\mu\right)&\simeq\hat{m}_s\left(\frac{\alpha_s}{\pi}\right)^{4 / 9}\left[1+0.895062\left(\frac{\alpha_s}{\pi}\right)+1.37143\left(\frac{\alpha_s}{\pi}\right)^2+1.95168\left(\frac{\alpha_s}{\pi}\right)^3\right] \\
%m_c\left(\mu\right)&\simeq\hat{m}_c\left(\frac{\alpha_s}{\pi}\right)^{12 / 25}\left[1+1.01413\left(\frac{\alpha_s}{\pi}\right)+1.38921\left(\frac{\alpha_s}{\pi}\right)^2+1.09054\left(\frac{\alpha_s}{\pi}\right)^3\right] \\
%%%%%%%%%%%%%%%%%%%%%%%%%%%%%
%Analogously, it is straightforward to obtain the results for the anomalous dimensions $\gamma_2,\gamma_3,\gamma_3^c$ related to the fields RCs (see e.g. Refs.\cite{Chetyrkin:2017bjc,Luthe:2017ttc}).

\section{The scheme-dependence  \label{rsdependence}}
\subsection{The $\Lambda$ parameter}

Being the values of the coefficients $\beta_0,\beta_1$ scheme invariant, the only parameter that can be fixed at NLO by the RS, is $\Lambda$. This parameter is also related to the position of the Landau ghost pole in perturbative QCD. A Landau pole was originally discovered in the QED coupling. However, the
presence of this pole doesn’t affect QED, being its value,
$\Lambda\sim10^{30-40}\,{\rm GeV}$, above the Planck scale
\cite{Gockeler:1997dn}, where new physics is expected to occur in order to restore the correct physical behavior. The QCD $\Lambda$ parameter in contrast characterises the low energy behavior of the strong coupling, its value depends on the RS, on the order of the $\beta$-series, $\beta_{i}$, on the approximation of the coupling $\alpha_s(\mu)$ at orders higher than $\beta_{1}$ and on the number of flavors~$N_f$. Although mass
corrections due to light quarks at higher order in perturbative
calculations introduce negligible terms, they actually indirectly
affect $\alpha_s$ through~$N_f$. In fact, the number of active
quark flavors runs with the scale $Q$ and a quark $q$ is
considered active in loop integration if the scale $Q\geq m_q$.
Thus, in general, light quarks can be considered massless
regardless of whether they are active or not, while $\alpha_s$
varies smoothly when passing a quark threshold, rather than in
discrete steps. The matching of the values of $\alpha_s$ below and
above a quark threshold makes $\Lambda$ depend on~$N_f$. Matching
requirements at leading order $\beta_0$, imply that:
\begin{align}
\alpha_{s}^{N_f-1}\left(Q{=}m_{q}\right)=
\alpha_{s}^{N_f}\left(Q{=}m_{q}\right)
\end{align}
and therefore that:
\begin{align}
\Lambda^{N_f}=\Lambda^{N_f-1}\left(\frac{\Lambda^{N_f-1}}{m_{q}}\right)^{2
/\left(33-2 N_f\right)}
\end{align}
The formula with $\beta_{1}$, can be found in \cite{Larin:1994va}
and the four-loop matching in the $\overline{\rm MS}$ RS is given
in \cite{Chetyrkin:1997sg}.
The value of $\Lambda$
is often associated with the confinement scale, or equivalently with
the hadronic mass scale. An explicit relation between hadron
masses and the $\Lambda$ scale has been obtained in the framework
of holographic QCD \cite{Brodsky:2014yha}. Landau poles on the
other hand, usually do not appear in nonperturbative approaches,
such as AdS/QCD. In general, one may think that lower values of the scale parameter lead to slower increasing couplings in the IR. Unfortunately other effects can occur spoiling this criterion. In fact the nature of
the perturbative expansion is affected by the renormalon
growth~\cite{tHooft1979} of the coefficients. 

\begin{table}
\begin{center}
\begin{tabular}{|c|c|c|c|c|}
\hline \multicolumn{5}{|c|}{ The numerical values of $\Lambda$ in
different schemes, ${\rm MeV}$}\\ \hline $N_{f}$ & the order of
approximation $\nu$ & $\Lambda_{\overline{\rm
MS}}^{\left(N_{f}\right)}$ & $\Lambda_{\rm
V}^{\left(N_{f}\right)}$ &
$\Lambda_{\rm mMOM}^{\left(N_{f}\right)}$ \\
\hline 4 & 2 & 350 & 500 & 625 \\
\hline 4 & 3 & 335 & 475 & 600 \\
\hline 4 & 4 & 330 & 470 & 590 \\
\hline 5 & 2 & 250 & 340 & 435 \\
\hline 5 & 3 & 245 & 335 & 430 \\
\hline 5 & 4 & 240 & 330 & 420 \\
\hline
\end{tabular}
\caption{Results for the $\Lambda$ parameter in different schemes,
at different values of the number of active flavor, $N_{f}$, and
at different orders of accuracy $\nu$~\cite{Kataev:2015yha}.}
\label{tablambda}
\end{center}
\end{table}

%%%%%%%%%%%%%%%%%%%%%%%%%%%%%%%%

%%%%%%%%%%%%%%%%%%%%%%%%
Different schemes can be related perturbatively by: \begin{align}
\alpha_{s}^{(2)}\left(Q\right)=\alpha_{s}^{(1)}\left(Q\right)\left[1+v_{1}
\alpha_{s}^{(1)}\left(Q\right) /(4 \pi)\right]
+\mathcal{O}(\alpha_s^2) \label{eq:schemet1}\end{align} 
where $v_{1}$ is the leading order
difference between $\alpha_{s}\left(Q \right)$ in the two schemes.
Eq.~\ref{eq:schemet1} can be obtained by considering the scale shift from
$\Lambda_{1}$ in the scheme 1 to $\Lambda_{2}$ in a scheme 2, which leads to the relation:
\begin{align}
\Lambda_{2}=\Lambda_{1} e^{\frac{v_{1}}{2 \beta_{0}}}.
\end{align}
Approximate values of $\Lambda$ in different
schemes, for different values of the number of active flavors $N_f$ and order of accuracy $\nu$, are given in Table \ref{tablambda}.
This relation is valid at each threshold, translating all values for
the scale from one scheme to the other. For further insights, relations among different RS and their associated $\Lambda$ are discussed in Refs.~\cite{Celmaster:1979km,Kataev:2015yha}.\\
%%%%%%%

\subsection{The $\beta_i$ coefficients in different schemes}
The scheme dependence at NNLO can be cast into the $\beta_{i}, i \geq2,\ldots$. In fact, though the first two coefficients $\beta_0,\beta_1$ are universally scheme-independent coefficients, depending only on the
number of colors $N_c$ and flavors $N_f$, the higher-order terms
are, in contrast, scheme dependent. In particular, for the 't
Hooft scheme \cite{tHooft1979} the higher $\beta_i, ~i\geq 2$
terms are set to zero, leading to the solution of
Eq.~\ref{lambert1} for the $\beta$-function valid at all orders.
Moreover, in all $\rm MS$-like schemes all the $\beta_{i}$
coefficients are gauge independent, while other schemes, such as
the momentum space subtraction (MOM) scheme
\cite{Celmaster:1979dm,Celmaster:1979km}, depend on the particular
gauge. Using the Landau gauge, the $\beta$ terms for the MOM
scheme are given by \cite{Boucaud:2005rm}
\begin{align}
\beta_{2}=3040.48-625.387 N_f +19.3833 N_f^{2}
\end{align}
and
\begin{align}
\beta_3=100541-24423.3 N_f+1625.4 N_f^{2}-27.493 N_f^{3}.
\end{align}
Results for the minimal MOM scheme and Landau gauge are given in
Ref.~\cite{Chetyrkin:2000dq}. The renormalisation condition for
the MOM scheme sets the virtual quark propagator to the same form
as a free massless propagator. Different MOM schemes exist and the
above values of $\beta_{2}$ and $\beta_3$ are determined with the
MOM scheme defined by subtracting the 3-gluon vertex to a point
with one null external momentum. This leads to a coupling that is
not only RS dependent but also gauge dependent. The values of
$\beta_{2}$ and $\beta_3$ given here are only valid in the Landau
gauge. Values in the $\mathrm{V}$-scheme defined by the static
heavy-quark potential \cite{Appelquist:1977tw,Fischler:1977yf,Peter:1996ig,Schroder:1998vy,Smirnov:2008pn,Smirnov:2009fh,Anzai:2009tm} can be found in Ref.~\cite{Kataev:2015yha}. They
result in $\beta_{2}=4224.18-746.01 N_f+20.88 N_f^{2}$ and
$\beta_3=43175-12952 N_f+707.0 N_f^{2}$ respectively. We recall
that the signs of the $\beta_{i}$ control the running
of~$\alpha_{s}$. We have $\beta_{0}>0$ for $N_f \leq 16,
\beta_{1}>0$ for $N_f \leq 8, \beta_{2}>0$ for $N_f \leq 5$ and
$\beta_3$ is always positive. Consequently, $\alpha_{s}$ decreases
at high momentum transfer, leading to the asymptotic freedom of
pQCD. Note that, $\beta_{i}$ are sometimes defined with an
additional multiplying factor $1 /(4 \pi)^{i+1}$.
 Different schemes are characterized by different $\beta_i, i\geq 2$ and lead to different definitions for the effective coupling.
 For a review of the strong coupling and of all the $\beta$ coefficients, see e.g.~\cite{Deur:2016tte,Blumlein:2023aso}.

\subsection{The extended renormalisation group}

Given that, physical predictions should not depend on the particular choice of the renormalisation scale nor on the choice of the scheme, one should include in the RGI also the invariance under scheme transformations. Thus, by using the same approach that led to the RG and its equations, it is possible to extend the group to scheme transformations. The wider group is also known as
{\it extended renormalisation group}. This was initially introduced by St\"uckelberg and
Peterman~\cite{StueckelbergdeBreidenbach:1952pwl}, then discussed
by
Stevenson~\cite{Stevenson:1980du,Stevenson:1981vj,Stevenson:1982wn,Stevenson:1982qw}
and also improved by Lu and Brodsky~\cite{Lu:1992nt}. A physical
quantity, $R$, calculated at the $N$-th order of accuracy is
expressed as a truncated expansion in terms of a coupling constant
$\alpha_{S}(\mu)$ defined in the scheme {\it Sc} and at the scale
$\mu$, such as: \begin{align}R_{N}=r_{0} \alpha_{Sc}^{p}(\mu)+r_{1}(\mu)
\alpha_{Sc}^{p+1}(\mu)+\cdots+r_{N}(\mu) \alpha_{Sc}^{p+N}(\mu).
\label{eqn:truncated}\end{align} At any finite order, the scale and scheme
dependencies of the coupling constant $\alpha_{S}(\mu)$ and of the
coefficient functions $r_{i}(\mu)$ do not totally cancel, this
leads to a residual dependence in the finite series and to the
scale and scheme ambiguities.

In order to generalize the RGE approach it is convenient to
improve the notation by introducing the universal coupling
function as the extension of an ordinary coupling constant to
include the dependence on the scheme parameters
$\left\{c_{i}\right\}:$

\begin{align}\alpha=\alpha\left(\mu / \Lambda,\left\{c_{i}\right\}\right) .
\end{align} where $\Lambda$ is the standard two-loop $\overline{\rm MS}$
scale parameter. The subtraction prescription is now characterized
by an infinite set of continuous {\it scheme parameters}
$\left\{c_{i}\right\}$ and by the renormalisation scale $\mu$.
Stevenson~\cite{Stevenson:1981vj} has shown that one can identify
the beta-function coefficients of a given renormalisation scheme
with the scheme parameters.  Considering that the first two
coefficients of the $\beta$-function are scheme independent, each
scheme is identified by its $\left\{\beta_{i}, \quad i=2,3,
\ldots\right\}$ parameters.

More conveniently, let us define the rescaled coupling constant
and the rescaled scale parameter as \begin{align}
a=\frac{\beta_{1}}{\beta_{0}} \frac{\alpha}{4 \pi}, \quad
\tau=\frac{2 \beta_{0}^{2}}{\beta_{1}} \log (\mu / \Lambda) . \end{align}
Then, the rescaled $\beta$-function takes the canonical form: \begin{align}
\beta(a)=\frac{d a}{d \tau}=-a^{2}\left(1+a+c_{2} a^{2}+c_{3}
a^{3}+\cdots\right) \end{align} with $c_{n}=\beta_{n} \beta_{0}^{n-1} /
\beta_{1}^{n}$ for $n=2,3, \cdots$.

The scheme and scale invariance of a given observable $R$, can be
 expressed as: \begin{align}\frac{\delta R}{\delta \tau} &= \beta
\frac{\partial
R}{\partial a}+\frac{\partial R}{\partial \tau}=0 \nonumber \\
\frac{\delta R}{\delta c_{n}}&=\beta_{(n)} \frac{\partial
R}{\partial a}+\frac{\partial R}{\partial c_{n}}=0.
\label{extendedrge} \end{align}

The fundamental beta function that appears in
Eqs.~\ref{extendedrge} reads: \begin{align}
\beta\left(a,\left\{c_{i}\right\}\right) \equiv \frac{\delta
a}{\delta \tau}=-a^{2}\left(1+a+c_{2} a^{2}+c_{3}
a^{3}+\cdots\right) \end{align} and the extended or scheme-parameter beta
functions are defined as: \begin{align}
\beta_{(n)}\left(a,\left\{c_{i}\right\}\right) \equiv \frac{\delta
a}{\delta c_{n}} . \end{align} The extended beta functions can be
expressed in terms of the fundamental beta function. Since the
$(\tau,\{c_i\})$ are independent variables, second partial
derivatives respect the commutativity relation: \begin{align}
\frac{\delta^{2} a}{\delta \tau \delta c_{n}}=\frac{\delta^{2}
a}{\delta c_{n} \delta \tau}, \end{align} which implies \begin{align}\frac{\delta
\beta_{(n)}}{\delta \tau}=\frac{\delta \beta}{\delta c_{n}},\end{align}
\begin{align}\beta \beta_{(n)}^{\prime}=\beta_{(n)} \beta^{\prime}-a^{n+2},
\end{align} where $\beta_{(n)}^{\prime}=\partial \beta_{(n)} / \partial a$
and $\beta^{\prime}=\partial \beta / \partial a$. From here \begin{align}
\beta^{-2}\left(\frac{\beta_{(n)}}{\beta}\right)^{\prime}=-a^{n+2},
\end{align} \begin{align}
\beta_{(n)}\left(a,\left\{c_{i}\right\}\right)=-\beta\left(a,\left\{c_{i}\right\}\right)
\int_{0}^{a} d x
\frac{x^{n+2}}{\beta^{2}\left(x,\left\{c_{i}\right\}\right)}, \end{align}
where the lower limit of the integral has been set to satisfy the
boundary condition
$$
\beta_{(n)} \sim O\left(a^{n+1}\right) .
$$
That is, a change in the scheme parameter $c_{n}$ can only affect
terms of order $a^{n+1}$ or higher in the evolution of the
universal coupling function.

The extended renormalisation group equations
Eqs.~\ref{extendedrge} can be written in the form: \begin{align}
\frac{\partial R}{\partial \tau}&= -\beta \frac{\partial R}{\partial a} \nonumber \\
\frac{\partial R}{\partial c_{n}}&=-\beta_{(n)} \frac{\partial
R}{\partial a}. \label{eqn:pms}\end{align}

Thus, provided we know the extended beta functions, we can
determine any variation of the expansion coefficients of $R$ under
scale-scheme transformations. In particular, we can evolve a given
perturbative series into another determining the expansion
coefficients of the latter and vice versa.
Thus, different schemes and scales can be related
according to the extended renormalisation group equations and the fundamental requirement of ``renormalisation scale and scheme invariance'' is recovered via the extended renormalisation group invariance of perturbative QCD.

\section{The renormalisation scale setting problem}

\subsection{The running coupling constant $\alpha_s(\mu)$ and the pQCD series \label{sec:alphas}}

%%%%%%

%\section{Conventional scale setting - CSS}

%According to common practice a first evaluation of the physical
%observable is obtained by calculating perturbative corrections in
%a given scheme (commonly used are $\rm MS$ or $\overline{\rm MS}$)
%and at an initial renormalisation scale $\mu_{r}=\mu_{r}^{\rm init
%}$,
In general, the QCD series for an observable is given by a truncated series in the strong coupling evaluated in a particular scheme {\it S } and at a particular scale, $\mu_{r}=\mu_{r}^{\rm init
}$. This can be written in the following form:

\begin{align}
\rho_{n}&=\mathcal{C}_{0}
\alpha_{s}^{p}\left(\mu_{r}\right)+\sum_{i=1}^{n}
\mathcal{C}_{i}\left(\mu_{r}\right)
\alpha_{s}^{p+i}\left(\mu_{r}\right), \quad(p \geq 0),
\label{observable-init0}
\end{align}
where $\mathcal{C}_{0}$ is the tree-level term, while
$\mathcal{C}_{1}, \mathcal{C}_{2}, ...,\mathcal{C}_{n}$ are the
one-loop , two-loop, n-loop corrections respectively and $p$ is
the power of the coupling at tree-level.

%
%In order to improve the pQCD estimate of the observable, after the initial renormalisation a change of scale using the RGE and a chosen scale-setting method is performed in
%Eq.~\ref{observable-init0}, which leads to: \begin{align}
%\rho_{n}&=\mathcal{C}_{0}
%\alpha_{s}^{p}\left(\tilde{\mu}_{r}^{0}\right)+\sum_{i=1}^{n}
%\overline{\mathcal{C}}_{i}\left(\tilde{\mu}_{\mathrm{r}}^{i}\right)
%\alpha_{s}^{p+i}\left(\bar{\mu}_{r}^{i}\right), \quad(p \geq 0)
%\label{observable-ren1} \end{align} where the new leading-order (LO) and
%higher-order scales $\tilde{\mu}_{r}^{0}$ and
%$\tilde{\mu}_{r}^{i}$ are functions of the initial renormalisation
%scale $\mu_{r}^{\rm init }$, and they depend the particular choice of the scale-setting method. At the same time, the new
%coefficients $\overline{\mathcal{C}}_{i}\left(\tilde{\mu}_{r}^{i}\right)$ are changed accordingly in order to obtain a consistent result. If in principle the
%Thus, theoretical predictions are affected by the actual values of the scheme and scale chosen in the QCD series.
%The simple CSS procedure starts from the scale and scheme
%invariance of a given observable, which translates into complete
%freedom for the choice of the renormalisation scale. In practice
%in this approach, the initial scale is directly fixed to the
%typical momentum transfer of the process , $Q$ , or to a value
%which minimizes the contributions of the loop diagrams and the
%errors are evaluated varying the value of $Q$ in the range of 2,
%$[Q / 2,2 Q]$.

%%%%%%

 The strong coupling $\alpha_s$, is the fundamental expansion parameter, the value of which depends on the renormalisation scale $\mu$ and on the renormalisation scheme. Thus, in order to give a thorough description of the hadronic interactions, it is necessary to determine the magnitude of the coupling and its behavior over a wide range of values, from low to high energy scales. Long and short distances are related to low and high energies respectively. In the high energy region the strong coupling has an {\it asymptotic behavior} and QCD becomes perturbative, while in the region of low energies, e.g. below the proton mass scale, the dynamics of QCD is affected by processes such as {\it quark confinement}, {\it soft radiation} and {\it hadronization}. In the first case experimental results can be matched with theoretical calculations and a precise determination of the $\alpha_s$ depends both on experimental accuracy and on theoretical errors. In the latter case experimental results are difficult to achieve and theoretical predictions are affected by the confinement and hadronization mechanisms, which are rather model dependent. Various processes also involve a precise knowledge of the coupling in both the high and low momentum transfer regions and in some cases calculations must be improved with electroweak (EW) corrections. Thus, the determination of the strong coupling over a wide range of energy scales is a crucial task in order to achieve results and to test QCD to the highest precision. Theoretical uncertainties in the value of
$\alpha_s(Q^2)$ contribute to the total theoretical uncertainty in
the physics investigated at the Large Hadron Collider (LHC), such as the Higgs sector, e.g. gluon fusion Higgs
production~\cite{Anastasiou:2016cez}. The behavior of the
perturbative coupling at low-momentum transfer is also fundamental for the scale of the proton mass, in order to understand hadronic structure, quark confinement and hadronization processes. Infrared (IR)
effects, such as {\it soft radiation} and {\it renormalon factorial growth}, spoil the perturbative nature of the QCD in the low-energy domain and thus its predictivity.
%%%%%%

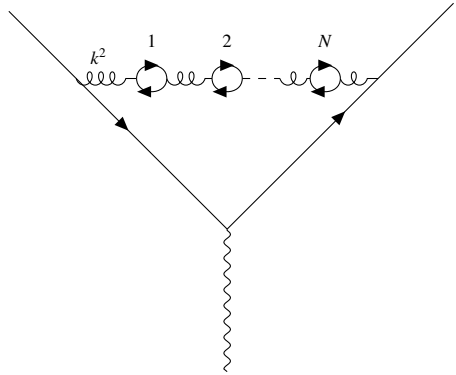
\begin{figure}[h]  
\begin{align}
& \begin{tikzpicture}
\begin{feynman}[small]
\vertex (a) at (0, 0);
\vertex  (d) at (3,3);
\node (e) at (-3,3);
\node (f) at (0,-2);
\vertex (b) at (-2,2);
\vertex (c) at (2,2);
\vertex  (v1) at (-1.2, 2);
\vertex (v2) at (-0.8,2);
\vertex  (v3) at (-0.2,2);
\vertex (v4) at (0.2,2);
\vertex  (v5) at (0.7,2);
\vertex  (v6) at (1.1,2);
\vertex  (v7) at (1.5,2);
\node  at (-1,2.5) {\scriptsize \bf $1$};
\node  at (0,2.5) {\scriptsize \bf $2$};
\node  at (1.3,2.5) {\scriptsize \bf $N$};
\node  at (-1.7,2.3) {\scriptsize \bf $k^2$};
\diagram*{
(a) -- [photon]  (f)  ;
(a)--[anti fermion] (e) ;
(a)--[fermion] (d) ;
(b)--[gluon] (v1) ;
(v1)--[fermion,  half left] (v2) ;
(v1)--[anti fermion,  half right] (v2) ;
(v2)--[gluon] (v3) ;
(v3)--[fermion,  half left] (v4) ;
(v3)--[anti fermion,  half right] (v4) ;
(v4)--[dashed] (v5) ;
(v5)--[gluon] (v6) ;
(v6)--[fermion,  half left] (v7) ;
(v6)--[anti fermion,  half right] (v7) ;
(v7)--[gluon] (c) ;
};  
\end{feynman}
\end{tikzpicture} \nonumber 
\end{align}
\caption{Example of a diagram with the ``bubble-chain" insertion.}\label{bubbles}
\end{figure}

The renormalons
affect renormalizable gauge theories only, they stem from
particular diagrams known as ``bubble-chain" diagrams and shown in
Fig.~\ref{bubbles}.

In loop integration these terms introduce a factorial growth,
using Eq.~\ref{1loopalphas} and considering $k^2 \geq \mu^2$ we
obtain:
$$\!\int \! d^4k k^{-2n}\alpha_s(k^2)\!=\! \alpha_s(\mu) \sum_N \int d^4k
k^{-2n} (\beta_0 \alpha_s(\mu) \ln(k^2/\mu^2))^N \!\sim\! \sum_N
N! (\frac{\beta_0}{n-2})^N \alpha_s(\mu)^{N+1},$$ with $n \geq 3$. Performing the Borel transform of the last series one obtains a geometric series, \begin{align}B(z)=\sum_N
\left(\frac{\beta_0}{n-2}\right)^N z^N
=\frac{1}{1-\left(\frac{\beta_0}{n-2}\right)z}\end{align} which has poles on the real positive axis: \begin{align}z_n=\frac{n-2}{\beta_0}, \quad
n=3,4,5,... \end{align} These singularities lead to an ambiguity in the inverse Borel transform given by the  non-zero residue contributions of
the type: \begin{align}\Delta=\left(\frac{\Lambda}{\mu}\right)^{2 \beta_0
z_n }. \end{align} IR and UV renormalons arise as singularities on the
real negative and positive axis of the complex $z$ plane of the
Borel transform, (analogously to instanton poles, this explains
the name given by 't Hooft~\cite{tHooft1979}), and are related to the $\Lambda$
scale and thus to the Landau pole of the strong coupling. These
terms affect the coefficients of the perturbative QCD series and
its convergence. It has been shown in several applications of
resummed quantities (i.e. applying the technique of resummation of
large IR logarithms) renormalons do not affect the final result if
one uses an appropriate prescription (e.g. the minimal
prescription (MP) formula~\cite{Catani:1989ne,DiGiustino:2011jn}).
Reviews on renormalons exist in the literature, e.g.
Refs.~\cite{Altarelli:1995kz,Beneke:1998ui}.

Thus, for a given observable different growths of the coefficients in different renormalisation schemes can balance the differences in values of the corresponding scales set by the choice of the scheme $\Lambda_{\rm sc}$. However, the growth of the coefficients is not only due to renormalons, in some cases the coefficients have an inherently fast rising behavior as shown in
Ref.~\cite{Brodsky:2000cr}.

%%%%%%%%%

Higher-twist effects can also play an important role. Processes
involving the production of heavy quarks near threshold require
the knowledge of the QCD coupling at very low momentum scales. Even reactions at high energies may involve the integration of the behavior of the strong coupling over a large domain of momentum scales including IR regions. Precision tests of the coupling are crucial also for other aspects of QCD that are still under continuous investigation, such as the hadron masses and their internal structure. In fact, the strong interaction is responsible for the mass of hadrons in the zero-mass limit of the u, d quarks.
%as in the case of the proton and the $\rho$ meson mass.

The origin and the phenomenology of the behavior of $\alpha_s(\mu)$ at short distances, where asymptotic freedom occurs, are well understood and explained in many textbooks on Quantum Field Theory and Particle Physics (see e.g. Refs.~\cite{Prosperi:2006hx,Altarelli:2013bpa}).

Other questions remain even in this well understood regime: a
significant issue is how to identify the scale Q that controls a
given hadronic process, especially when the process depends on many physical scales.

In fact, in the perturbative regime, theoretical predictions are affected by several sources of errors, e.g. the top and Higgs mass uncertainty, the strong coupling uncertainty and by the {\it missing higher orders} (MHO). The latter are also retained to be the main source for the errors related to  {\it renormalisation scale and scheme ambiguities}.

\subsection{The renormalisation scale and scheme ambiguities}

The scale-scheme ambiguities prevent
precise theoretical predictions for both SM and BSM physics. In
principle, an infinite perturbative series is void of this issue,
given the scheme and scale invariance of the entire physical
quantities~\cite{StueckelbergdeBreidenbach:1952pwl,GellMann:1954fq,Peterman:1978tb,Callan:1970yg,Symanzik:1971vw},
in practice perturbative corrections are known up to a certain
order of accuracy and scale invariance is only approximated in
truncated series, leading to the scheme and scale
ambiguities~\cite{Celmaster:1979km,Buras:1979yt,Stevenson:1980du,Stevenson:1981vj,Stevenson:1982qw,Stevenson:1982wn,Grunberg:1980ja,Grunberg:1982fw,Grunberg:1989xf,Brodsky:1982gc,Chishtie:2015lwk,Chishtie:2016wob,Abbott:1980hwa}.
If on one hand, according to the conventional practice, or
conventional scale setting (CSS), this problem cannot be avoided
and is responsible for part of the theoretical errors, on the
other hand some strategies for the optimisation of the truncated
expansion have been proposed, such as the Principle of Minimal
Sensitivity proposed by Stevenson~\cite{Stevenson:1981vj}, the
Fastest Apparent Convergence criterion introduced by
Grunberg~\cite{Grunberg:1980ja} and the Principle of Maximum Conformality (PMC)~\cite{Brodsky:2011zza,Brodsky:2011ig} which generalises the previous Brodsky-Lepage-Mackenzie method (BLM)~\cite{Brodsky:1982gc}. These are procedures commonly in use for scale setting in perturbative QCD. In general, a scale-setting procedure is considered reliable if it preserves important self consistency requirements. All Renormalisation Group
properties such as: {\it uniqueness, reflexivity, symmetry,} and
{\it transitivity} should be preserved also by the scale-setting
procedure in order to be generally applied~\cite{Brodsky:2012ms}.

Thus, in order to improve the pQCD estimate of the observable, after the initial renormalisation a change of scale using the RGE and a chosen scale-setting method is performed in
Eq.~\ref{observable-init0}, which leads to: \begin{align}
\rho_{n}&=\mathcal{C}_{0}
\alpha_{s}^{p}\left(\tilde{\mu}_{r}^{0}\right)+\sum_{i=1}^{n}
\overline{\mathcal{C}}_{i}\left(\tilde{\mu}_{\mathrm{r}}^{i}\right)
\alpha_{s}^{p+i}\left(\bar{\mu}_{r}^{i}\right), \quad(p \geq 0)
\label{observable-ren1} \end{align} where the new leading-order (LO) and
higher-order scales $\tilde{\mu}_{r}^{0}$ and
$\tilde{\mu}_{r}^{i}$ are functions of the initial renormalisation
scale $\mu_{r}^{\rm init }$, and they depend the particular choice of the scale-setting method. At the same time, the new
coefficients $\overline{\mathcal{C}}_{i}\left(\tilde{\mu}_{r}^{i}\right)$ are changed accordingly in order to obtain a consistent result. 
Given the fixed $p$ order of the accuracy in the series, it follows that theoretical predictions are affected by the actual values of the scheme and scale chosen. This problem only marginally occurs in QED, given the small value of the QED coupling and the perturbative nature of the theory up to very high values of the scale. In the QCD the coupling strength is much larger and thus the truncated expansion is severely affected by this issue. 
According to the CSS, given the RGI, there should be a complete freedom in choosing the scale and the scheme. However, this approach heavily relies on the convergence of the perturbative series, which is strictly process dependent and at large accuracies is actually asymptotic. 

Thus, the choice of a correct scale might solve or at least improve the results. We introduce in this section the different optimisation procedures and their properties. An introduction to these methods can also be
found in Refs.~\cite{Wu:2013ei,Deur:2016tte,DiGiustino:2023jiq}.

\subsection{Optimisation procedures}

\subsubsection{The Principle of Minimal Sensitivity: PMS Scale-Setting}

The Principle of Minimal
Sensitivity~\cite{Stevenson:1981vj,Stevenson:1982qw} derives from
the assumption that, since observables should be independent of
the particular RS and scale, their optimal perturbative
approximations should be stable under small RS variations. The RS
scheme parameters $\beta_2,\beta_3,...$ and the scale parameter
$\Lambda$ (or the
 subtraction point $\mu_r$), are considered as ``unphysical" and
independent variables, and then their values are set in order to
minimize the sensitivity of the estimate to their small
variations. This is essentially the core of the Optimized
Perturbation Theory (OPT)~\cite{Stevenson:1981vj}, based on the
PMS procedure. The convergence of the perturbative expansion,
Eq.~\ref{observable-init0}, truncated to a given order $\rho_n$,
is improved by requiring its independence from the choice of RS
and $\mu$. The optimisation is implemented by identifying the
RS-dependent parameters in the $\rho_n$-truncated series (the
$\beta_{i}$ for $2 \leq i \leq n$ and $\Lambda$ ), with the
request that the partial derivative of the perturbative expansion
of the observable with respect to the RS-dependent and scale
parameters vanishes. In practice the PMS scale setting is designed
to eliminate the remaining renormalisation and scheme dependence
in the truncated expansions of the perturbative series.

More explicitly, the PMS requires the truncated series, i.e. the
approximant of a physical observable defined in
Eq.~\ref{observable-init0}, to satisfy the RG invariance given by
the \ref{eqn:pms}, with the substitution of the proper
$\beta_{(n)}$ function:

 \begin{align}
  \frac{\partial
\alpha_{\mathrm{s}}}{\partial
\beta_{j}}&=-\beta\left(\alpha_{\mathrm{s}}\right)
\int_{0}^{\alpha_{\mathrm{s}}} \mathrm{d} \alpha^{\prime}
\frac{\alpha^{\prime
j+2}}{\left[\beta\left(\alpha^{\prime}\right)\right]^{2}}=\frac{\alpha_{\mathrm{s}}^{j+1}}{\beta_{0}}\left(\frac{1}{j-1}-\frac{\beta_{1}}{\beta_{0}}
\frac{j-2}{j(j-1)} \alpha_{\mathrm{s}}+\ldots\right) , \end{align} 
it follows that:

 \begin{align}
\frac{\partial \rho_{n}}{\partial
\tau}&=\left(\left.\frac{\partial}{\partial
\tau}\right.+\beta\left(\alpha_{s}\right) \frac{\partial}{\partial
\alpha_{s}}\right) \rho_{n}  \equiv 0 \\
 \frac{\partial \rho_{n}}{\partial
\beta_{j}}&=\left(\left.\frac{\partial}{\partial
\beta_{j}}\right.-\beta\left(\alpha_{\mathrm{s}}\right)
\int_{0}^{\alpha_{\mathrm{s}}} \mathrm{d} \alpha^{\prime}
\frac{\alpha^{\prime
j+2}}{\left[\beta\left(\alpha^{\prime}\right)\right]^{2}}
\frac{\partial}{\partial \alpha_{\mathrm{s}}}\right) \rho_{n}
\equiv 0
 \end{align} 
 where $\tau=\ln \left(\mu_{r}^{2} / \Lambda_{\rm
QCD}^{2}\right)$ and $j \geq 2$. Scheme labels have been omitted.
The request of RS-independence modifies the series coefficients
$\mathcal{C}_{i}\left(1 \leq i \leq n\right)$ and the coupling
$\alpha_{s}$ to the PMS ``optimised" values
$\widetilde{\mathcal{C}_{i}}$ and $\widetilde{\alpha_{s}}$. 
This procedure can be extended to higher order and it can be generally applied to calculations obtained in an arbitrary initial renormalisation scheme.

%%%%%%%%%%%%%%%%%%%%%%%%%%%%%%%%%%%%%%%%%%%%%%%%%%%%%%%%%%%%%%%%%%%%%%%%

\subsubsection{The Fastest Apparent Convergence principle - FAC scale setting}

%%%%%%%%%%%%%%%%%%%%%%%%%%%

The Fastest Apparent Convergence (FAC) principle is based on the
idea of {\it effective charges}. As pointed out by
Grunberg~\cite{Grunberg:1980ja,Grunberg:1982fw,Grunberg:1989xf},
any physical quantity can be used to define an effective charge, by entirely incorporating the radiative corrections into its definition.
Effective charges can be defined from an observable starting from the assumption that the infinite series of a given quantity is
scheme and scale invariant. Given the perturbative series
$R=\mathcal{C}_0 \alpha_s^{p}+\cdots,$ the relative effective
charge $\alpha_{R}$ is given by 
\begin{align}
 \alpha_{R} \equiv
\left(\frac{R}{\mathcal{C}_{0}}\right)^{1/p}. 
\end{align} Since $R,
\mathcal{C}_{0}$ and $p$ are all renormalisation scale and scheme invariant, the effective charge $\alpha_{R}$ is scale and scheme invariant.

The effective charge satisfies the same renormalisation group
equations as the usual coupling. Thus, the running behavior for
both the effective coupling and the usual coupling are the same if
their RG equations are calculated in the same renormalisation
scheme. This idea has been discussed in more detail in
Refs.~\cite{Dhar:1983py,Gupta:1990jq}.

Using the effective charge $\alpha_{R}$, the ratio $R_{\rm
e^+e^-}$ becomes~\cite{Gorishnii:1990vf}:
\begin{align}
R_{e^{+} e^{-}}\left(Q^{2}\right) & \equiv R_{e^{+}
e^{-}}^{0}\left(Q^{2}\right)\left[1+\frac{\alpha_{R}(Q)}{\pi}\right]
\end{align}
where $R_{e^{+} e^{-}}^{0}\left(Q^{2}\right)$ is the Born result
and $s=Q^{2}$ is the center-of-mass energy squared.

We notice that all effective couplings defined in the same scheme satisfy the same RG equations. While different
schemes or effective couplings, will differ through the third and
higher coefficients of the
$\left\{\beta_{i}^{\mathcal{R}}\right\}$-functions, which are
scheme $\mathcal{R}$ dependent. Hence, any effective coupling can be used as a reference to define the renormalisation procedure.
Given that expansions of the effective charges are known only up to a certain order, $\alpha_{R} \simeq
\left(\frac{R_n}{\mathcal{C}_{0}}\right)^{1/p}$, an optimisation
procedure is used to improve the perturbative calculations, namely the FAC scale setting. The basic idea of the FAC scale setting method is to set to zero all the higher order perturbative
coefficients, i.e. $\mathcal{C}_{i(\geq 1)}\left(\mu_{r}^{\rm
FAC}\right) \equiv 0$, including all fixed order corrections into
the FAC renormalisation scale of the leading term by means of the RG equations in order to provide a reliable
estimate~\cite{Krasnikov:1981rp}. In general this method can be applied to any observable calculated in any RS and at any order of accuracy.

\subsubsection{The Principle of Maximum Conformality -- PMC scale
setting}\label{sec:blm}

\begin{figure*}[htb]
\includegraphics[width=8cm,height=6cm]{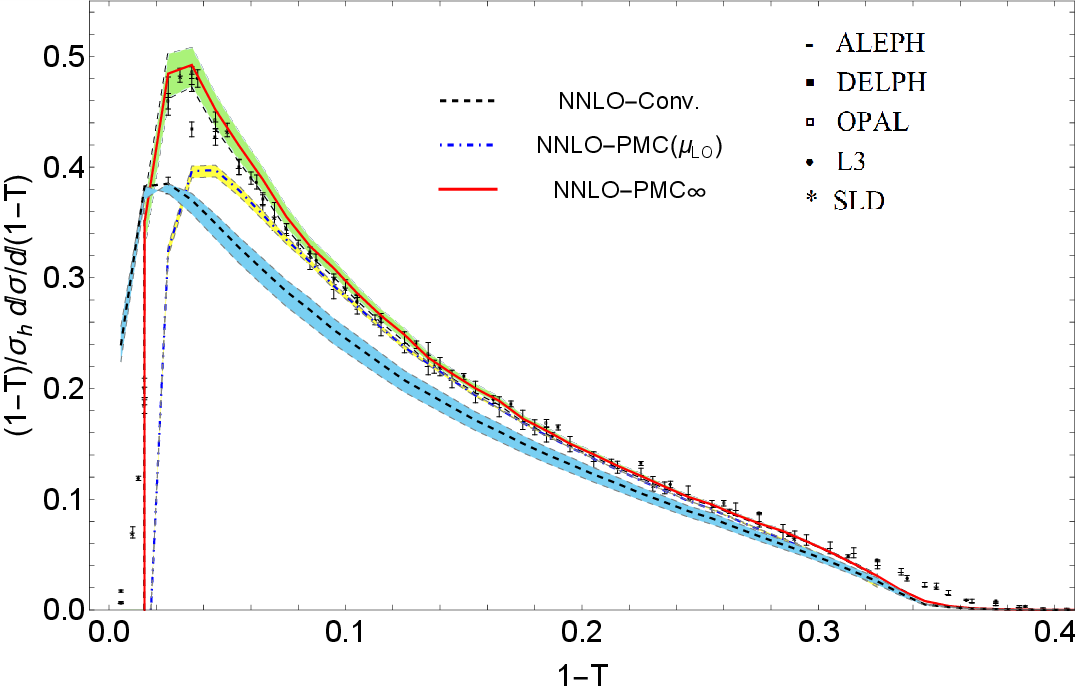}
\includegraphics[width=8cm,height=6cm]{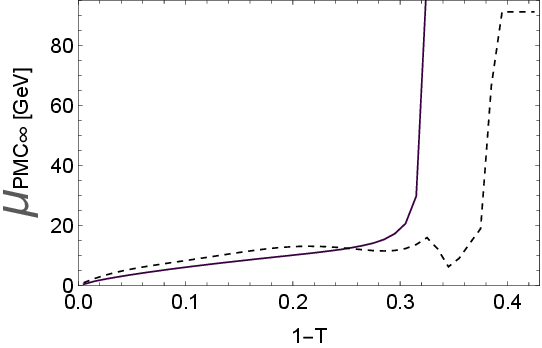}
  \caption{On the left, the thrust distribution at NNLO under the Conventional (dashed black), the PMC($\mu$\textsubscript{LO}) (dotdashed Blue) and the PMC$_\infty$ (solid red). The experimental data points are
taken from the ALEPH, DELPHI,OPAL, L3, SLD experiments
\cite{ALEPH:2003obs,DELPHI:2003yqh,OPAL:2004wof,L3:2004cdh,SLD:1994idb}. The shaded areas show
theoretical errors predictions at NNLO and they have been
calculated varying the remaining initial scale value in the range
$\sqrt{s}/2 \leq \mu_0 \leq 2 \sqrt{s}$~\cite{DiGiustino:2020fbk}. On the right,The LO-PMC$_\infty$ (solid black) and the NLO-PMC$_\infty$ (dashed black) scales for thrust~\cite{DiGiustino:2020fbk}. }
  \label{fig:one}
\end{figure*}

\begin{figure} [htb]
\centering
\includegraphics[width=8.5cm]{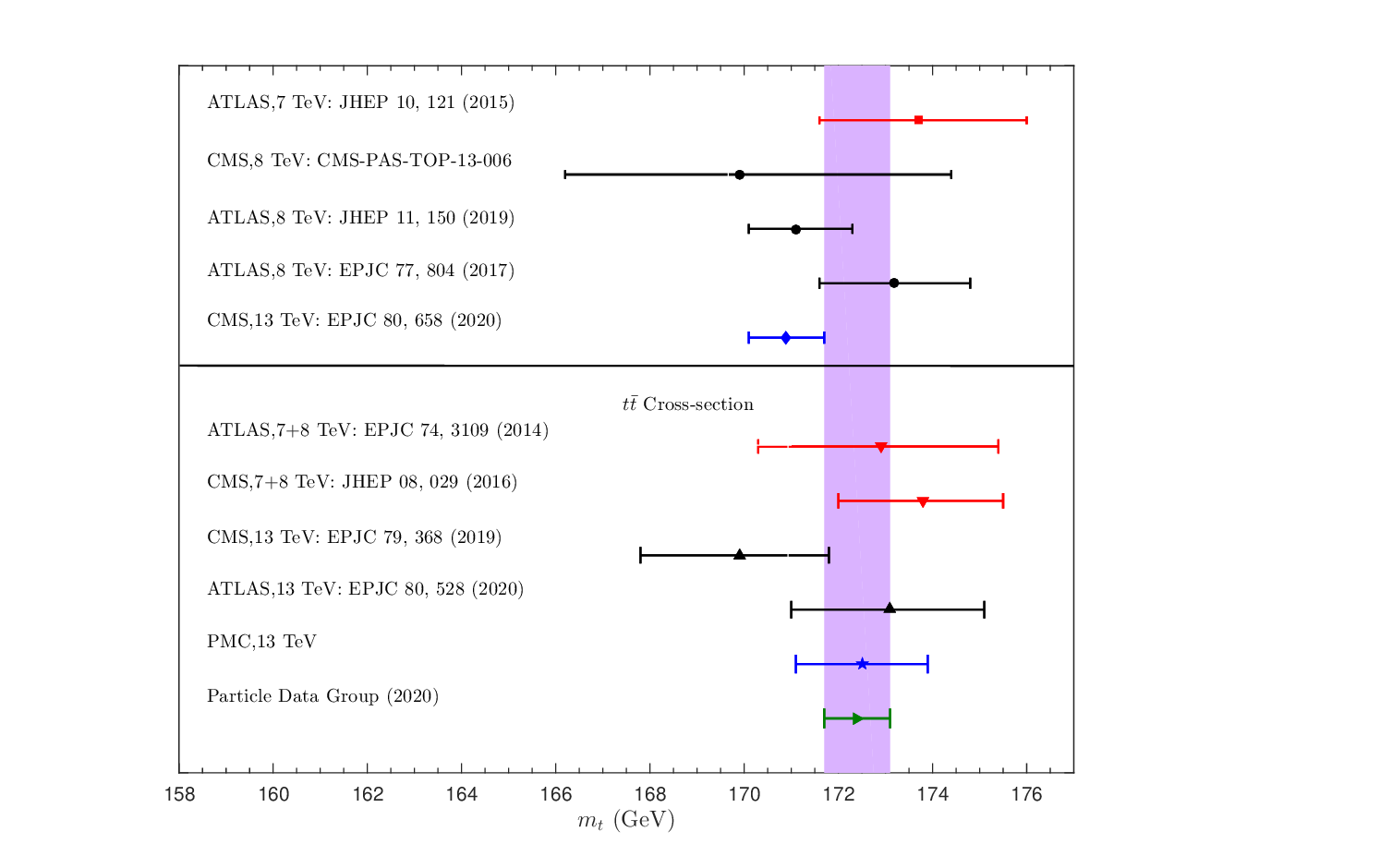} 
 \includegraphics[width=8cm]{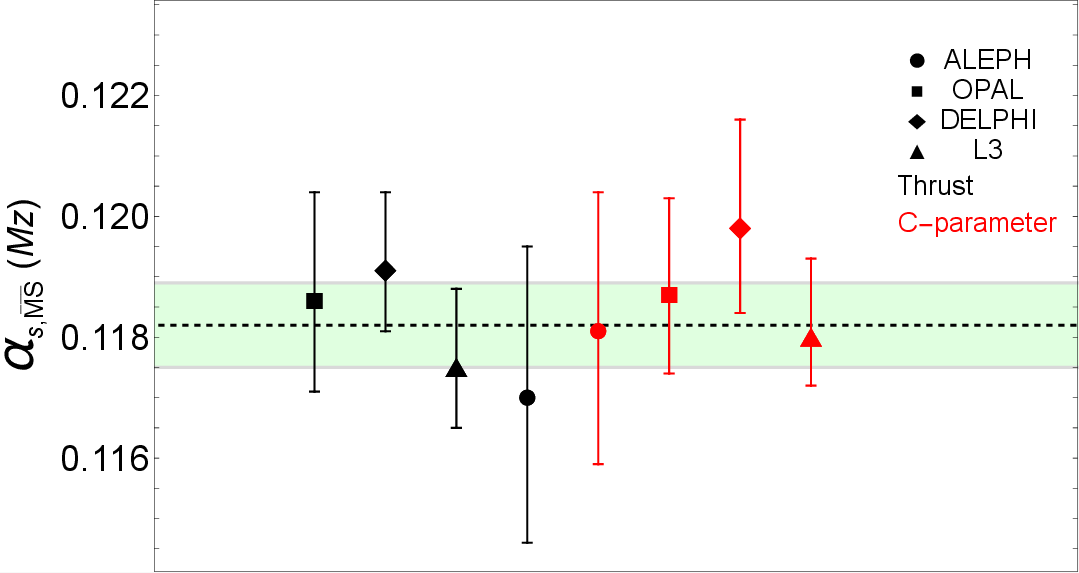}
\caption{ On the left, summary of the top-quark pole masses,
 where the PMC result and previous determinations from collider measurements at different energies and different techniques are presented.
 The top-quark pole mass from the PDG~\cite{ParticleDataGroup:2020ssz} is also presented as the shaded band for reference~\cite{DiGiustino:2023jiq}. On the right, Results for the strong coupling $\alpha_s(\mu)$ at the $Z^0$ peak for thrust and
  $C$-parameter, using the PMC$_\infty$ method and the data from ALEPH, OPAL, DELPHI and L3 \cite{ALEPH:2003obs,OPAL:2004wof,DELPHI:2003yqh,L3:2004cdh}.
The shaded area shows the total errors for the average value $\alpha_s(M_Z)= 0.1182^{+0.0007}_{-0.0007}$~\cite{DiGiustino:2024zss}.}
\label{fig:two}
\end{figure}

The Principle of Maximum Conformality
(PMC) \cite{Brodsky:2011ig, Brodsky:2011ta, Brodsky:2012rj, Mojaza:2012mf, Brodsky:2013vpa}
is the principle underlying BLM and it generalises the BLM method to all possible applications and to all orders.
Starting from a reparameterization of Eq.\ref{observable-init0}, as follows:

\begin{align}
\rho(Q)&=r_{1,0}a_s(Q) + \bigg(r_{2,0}+\beta_{0}r_{2,1}\bigg)\, a_{s}^{2}(Q)\nonumber
+\bigg(r_{3,0}+\beta_{1}r_{2,1}+ 2\beta_{0}r_{3,1}+ \beta_{0}^{2}r_{3,2}\bigg)\, a_{s}^{3}\, (Q)\nonumber \\
 &+\bigg(r_{4,0}+\beta_{2}r_{2,1}+ 2\beta_{1}r_{3,1} + \frac{5}{2}\beta_{1}\beta_{0}r_{3,2} 
+3\beta_{0}r_{4,1}+3\beta_{0}^{2}r_{4,2}+\beta_{0}^{3}r_{4,3}\bigg)\, a_{s}^{4}(Q)+\cdots, ~\label{rij1}
\end{align}
where $r_{i,j}$ can be derived from $\mathcal{C}_{i}$ by isolating the $N_f$-terms and transforming them into $\beta$-terms; the $r_{i,0}$ coefficients do not depend on the scale, i.e. they are {\it conformal}, while the $r_{i,j}$ $(j\neq0)$ are {\it non-conformal}.
By a shift of the scale at each order it is possible to reabsorb the non-conformal coefficients into the PMC-scales via the RGE.
This procedure leads to the {\it conformal series}:
\begin{align}
\rho(Q)=\sum_{n} r_{n,0}\, a^n_s(\mu^{(n)}_{\rm PMC})
~\label{eq:conformalseries}
\end{align}
where the scales $\mu^{(n)}_{\rm PMC}$, $n=1,2,3\ldots$ are set by $\beta$-terms at each order.
These terms stem directly from the loop-integration and they reveal the subprocess quark and gluon QCD dynamics, which determines the strength of the running coupling. As shown in the Fig.~\ref{fig:one} (right) the PMC scales are not necessarily single-valued functions, but they rather depend on the unintegrated variables. This allows to a determination of the strong coupling over a wider range of energies from a single experiment, as shown in Refs. \cite{Wang:2019isi,Wang:2021tak}. Fundamental result of this method is the elimination of the scale and scheme ambiguities that plague the perturbative predictions, leading to a significant reduction of the theoretical errors (see e.g. Ref.~ \cite{DiGiustino:2023jiq}). The PMC preserves all fundamental properties of the RG, such as \emph{reflexivity},
\emph{symmetry} and \emph{transitivity}. 
In fact, by this method, it is possible to relate couplings in different schemes using the \emph{commensurate scale relations} (CSRs)~\cite{Lu:1991yu, Lu:1991qr, Brodsky:1995ds, Brodsky:1994eh}. Using this approach it is also possible to extend conformal properties to renormalizable gauge theories, e.g. the generalised Crewther relation \cite{Crewther:1972kn,Broadhurst:1993ru,Baikov:2010je,Brodsky:1995tb}.
The PMC reduces to the GM--L scheme for QED in the Abelian
limit $N_c \rightarrow 0$ \cite{Brodsky:1997jk} leading to results that agree with QED.
Given that the $\beta$ terms are reabsorbed into the PMC scales, renormalon terms, $n!\beta^n_0\alpha^{n+1}_s$, cancel over the entire range of the accessible physical energies. Features of this method are shown in Refs. \cite{Brodsky:2011ta, Brodsky:2012rj, Brodsky:2011ig, Mojaza:2012mf, Brodsky:2013vpa}. More recent formalisations are discussed in detail in Refs. \cite{Huang:2021hzr,DiGiustino:2020fbk}. So far, the PMC has been applied to several fundamental processes where the renormalisation scale plays an important role in the precision of the theoretical predictions (see e.g.  Refs.~\cite{Ren:2025uns,Wang:2025irh,Wang:2025afy,Wang:2020ckr}).
Results for the comparison between the conventional scale setting and PMC method for thrust in $e^+e^-$ annihilation process, are shown in Fig.\ref{fig:one} (left) . Determination of the strong coupling at the $Z^0$-mass from thrust and C-parameter and the determination of the top-mass using the PMC, are shown in  Fig. \ref{fig:two} (right) and (left) respectively.
This method stems from first principles and in general it is applicable to any renormalizable gauge theory, to any process and to all orders. The PMC offers the possibility to use the same procedure to set the renormalisation scale to the entire SM, including the Yukawa sector. This is a crucial task for a scale-setting procedure in the perspective of a theory unifying all forces, such as the so-called {\it grand unified theory} (GUT), which constrains to the use of a single procedure in all sectors.

%%%%

\section{Conclusions}
\label{sec:conclusions}

In this chapter, we have introduced basic concepts and formalism of the renormalisation techniques. Theoretical foundations have been discussed, giving also the reader suggestions for further readings. This chapter is more intended for beginning graduate students that want to enter quickly into renormalisation and may serve as a reference for possible investigations or insights into the subject.
We have introduced the renormalisation scale setting problem in QCD and the current state of the art in scale setting procedures, then we have shown recent developments in applications to perturbative QCD calculations.
In the present phase of the LHC experiment and in the perspective of higher precision future colliders, the scale setting problem would be an obstacle for precise theoretical predictions. Thus, the use of a proper procedure would be crucial in order to improve the precision in perturbative calculations.

\begin{ack}[Acknowledgments]%
The author thanks Philip G. Ratcliffe for helpful comments.
\end{ack}

%%%%%%%%%%%%%%%%%%%%%%%%%%%%%%%%%%%%%%%%%%%%
%% Optional: A list of references to other relevant works/articles/websites which are not cited in the text but that would further enhance a readers understanding of this topic
%\seealso{article title article title}

%%%%%%%%%%%%%%%%%%%%%%%%%%%%%%%%%%%%%%%%%
%% Mandatory: Bibliography using bibtex 
\bibliographystyle{Numbered-Style} %% for Numbered Reference Style
\bibliography{bibliopedia}

\end{document}